\documentclass[11pt,a4paper]{article}
\usepackage[authoryear]{natbib}
\usepackage{graphics}
\usepackage{ifpdf}
\usepackage{graphicx}
\usepackage{amssymb,amstext}
\usepackage{amsmath,latexsym,xspace}
\usepackage{array}
\usepackage{setspace} 
\usepackage{comment}
\usepackage{authblk}
\usepackage{diagbox}
\usepackage{multirow}
\usepackage{caption}
\usepackage{subcaption}
\usepackage{hyperref}

\usepackage{amsthm}

\usepackage[scale={0.75,0.85},centering,includeheadfoot]{geometry}

\def\mm#1{\ensuremath{\boldsymbol{#1}}} 
\def\AR1{AR$(1)$\xspace}

\setkeys{Gin}{width=0.45\textwidth}

\title{Careful prior specification avoids incautious inference for log-Gaussian Cox point processes}
\author[1]{Sigrunn H. S{\o}rbye}
\author[2]{Janine B. Illian}
\author[3]{Daniel P. Simpson}
\author[4]{David Burslem}
\affil[1]{Department of Mathematics and Statistics, UiT The Arctic University of Norway, Troms{\o}, Norway}
\affil[2]{Centre for Research into Ecological and Environmental Modelling, University of St. Andrews,  UK} 
\affil[3]{Department of Statistical Sciences, University of Toronto, Canada}
\affil[4]{School of Biological Sciences, University of Aberdeen, UK}

\begin{document}
\maketitle

\begin{abstract}
    Prior specifications for hyperparameters of random fields in
    Bayesian spatial point process modelling can have a major impact
    on the statistical inference and the conclusions made. We consider
    fitting of log-Gaussian Cox processes to  spatial point patterns
    relative to spatial covariate data. From an ecological point of
    view, an important aim of the analysis is to assess significant
    associations between the covariates and the point pattern
    intensity of a given species. This paper introduces the use of a
    reparameterised model to facilitate meaningful interpretations of
    the results and how these depend on hyperprior specifications. The
    model combines a scaled spatially structured field with an
    unstructured random field, having a common precision parameter. An
    additional hyperparameter identifies the fraction of variance
    explained by the spatially structured term and proper scaling
    makes the analysis invariant to grid resolution. The
    hyperparameters are assigned penalised complexity priors, which
    can be tuned intuitively by user-defined scaling parameters. We
    illustrate the approach analysing covariate effects on point
    patterns formed by two rainforest tree species in a study plot on
    Barro Colorado Island, Panama.
\end{abstract}

{\bf Keywords:} Bayesian inference, ecological point process, penalised complexity prior, \texttt{R-INLA}, spatial modelling

\section{Introduction}\label{sec:intro}
\subsection{Habitat association modelling}
There is a strong ecological interest in understanding the mechanisms that allow different species to coexist within an ecosystem, and in particular in extremely species-rich systems, such as rainforests or coral reefs \citep{chesson2000mechanisms, hubbell:01, dornelas2006coral}. Habitat association has often been seen as a potential strategy that reduces interspecific competition and hence facilitates coexistence \citep{connell1961influence, bagchi2011spatial}.
Ecological research has highlighted associations of specific plant species with distinct  habitat defined by, for instance, climate and soil, but in many cases these associations have been derived from anecdotal knowledge expressed in floras rather than from a study that has thoroughly tested them. 
Recently, increasingly detailed data on soil properties have become available which provide fine-scale, spatially continuous data on soil chemistry rather than discrete habitat types providing an opportunity for a more detailed and local analysis  characterising a species' environmental preferences. 

A statistical analysis of the relevant data structure here  would need to relate the individuals' occurrence in a specific location to the local environment. Spatial point process methodology  \citep{diggle:03, illian:08, baddeley:15}  has been used in this context \citep{law2009ecological, wiegand:14,velazquez:16} as it models the pattern formed by individuals in space relative to local conditions.
Common approaches relate the occurrence of an individual in a location in continuous space to a set of environmental covariates. In addition to the terms representing these spatial covariates and associated parameters, models contain a further term, for instance a spatially structured random field such as a Gaussian random field,  which accounts for spatial structures unexplained by the covariates \citep{illianal-a:12, waagepetersen:16}. These structures might be the result of biotic mechanisms such as dispersal limitation or of unobserved covariates. 

However,  this modelling approach is not as straight forward as it might seem. In particular,  the literature is lacking a thorough discussion of (a) the impact of the smoothness of the spatial field on inference and (b) complexity of accounting for unexplained spatial structures due to mechanisms operating at a number of spatial scales. With regard to (a), the level of significance of the covariates' association with the observed spatial pattern in a fitted model is not independent of the smoothness of the spatially structured field, and determining this smoothness in an objective way is difficult, see Section  \ref{statspersp}. Furthermore, the remaining spatial structure is likely to operate at different spatial scales, e.g.\  as a combination of large scale clustering due to habitat association and small scale clustering due to local dispersal, and these are difficult to disentangle \citep{hamill1986testing, bagchi2011spatial}.
In both the statistical literature on point processes and the ecological literature using spatial point process methods these issues have not been explicitly resolved. To the best of our knowledge a discussion of principles for choosing the smoothness of the spatial field is entirely absent from the spatial point process literature with a few papers accounting for the issue of clustering at different spatial scales \citep{john:07, illianal-a:12, shen2013quantifying}.

In this paper, we acknowledge these issues and that there is no universal objective criterion for resolving them.  We discuss Bayesian methodology that  allows us to investigate the spatial behaviour by a) constructing a model that explicitly takes the awareness of these issues into account and b) suggesting a constructive way of addressing them. This is done by using interpretable priors that allow us to vary the focus on different spatial scales  and to observe changes in the resulting estimates with changes in scale. In other words we do not attempt to identify a single degree of smoothing that is optimal by some criterion but consider a standardised range of degrees of smoothing and observe the behaviour of the estimates relative to changes in smoothness.

The approach discussed here has been motivated by the analysis of point patterns formed by tropical forest tree species within the 50-ha rectangular plot at Barro Colorado Island (BCI), Panama. This plot was designed to help address the mechanisms that maintain species richness and deliberately involved spatial mapping of the entire community following the recognition that population and community dynamics occur in a spatial context  \citep{hubbell:01}.
The full
dataset includes observed positions for a large number of tree species
\citep{condit:98, hubbell:99, hubbell:05} and measurements of
topographical variables and soil nutrients that potentially influence
the spatial distribution of the trees \citep{john:07, schreeg:10}. A
main aim of the statistical analysis is to investigate whether the
spatial patterns of different tree species can be related to spatial
environmental variations, reflected by observed topography and
soil nutrients \citep{moeller:07}. To appropriately inform model choice and inference, this paper suggests an approach that does neither attempt to universally suggest   an optimal degree of smoothing nor to a fixed spatial scale at which mechanisms are operating. This is done by explicitly communicating prior choices that impact on inference and statistical conclusions. 

\subsection{The statistical perspective}\label{statspersp}

Point pattern data are the observed locations of
objects or events within a bounded geographical region, for example
the locations of plants or animals in the wild or the locations of
earthquakes. By fitting a spatial point process model, the spatial
structure of the pattern can be studied in terms of observed
environmental variables that are included as
covariates. To perform reliable statistical inference and to make an
honest accounting of estimation uncertainty, it is of vital importance
to also account for spatial dependence structures and random error due
to potentially missing covariates and biotic processes. The resulting
model will be sensitive to input choices. Hence, extra effort must be
made to correctly communicate statistical statements and conclusions.

This paper explicitly focuses on how hyperprior
specifications influence the statistical inference when a discretized
log-Gaussian Cox process \citep{moeller:98}  is fitted to an observed
spatial point pattern, including covariate information. However, all of the conclusions in this paper  remain valid for any point process model that includes a high- or infinite-dimensional parameter as well as covariates. A log-Gaussian
Cox process is a Poisson process with random intensity
$\Lambda(s)=\exp\{\eta(s)\},$ where $\{\eta(s): s\in \mathbb{R}^2 \}$
is a latent Gaussian random field. To assess the influence of observed
covariates, we can express the log-intensity $\eta(s)$ in a spatial
grid cell $s$ as
\begin{equation}
    \eta(s) = \beta_0+\sum_{j=1}^{n_{\beta}}\beta_j z_{j}(s) +u(s) +v(s), \quad s\in \Omega, \label{eq:model-old}
\end{equation}
where $\Omega$ represents a bounded two-dimensional study region.
Here, $\beta_0$ is an intercept while $\{\beta_j\}_{j=1}^{n_{\beta}}$ represent linear
fixed effects of observed covariates $\{z_j(s), s\in \Omega\}_{j=1}^{n_{\beta}}$. The
spatially-correlated random field $\mm{u}=\{u(s),\,s\in \Omega$\} is included to account for
spatial autocorrelation or over-dispersion in point counts among
neighbouring grid cells.
The random field $\mm{v}=\{v(s),\,s\in \Omega$\} is referred to as an error field or a
spatially unstructured effect, accounting for over-dispersion or
clustering within grid cells. A common modelling strategy is to assign
Gamma priors to the precision (inverse variance)  parameters of the two
random fields \citep{rueal:09, illianal-a:12, illianal-b:12, kang:15}.
However, both the model formulation in (\ref{eq:model-old}) and the
use of Gamma priors are burdened with problems:
\begin{enumerate}
\item The two random fields in the model are not independent as
    $\mm{v}$ can be seen to be included in $\mm{u}$ in situations with
    no spatial dependence. The priors for the precision parameters
    should therefore not be chosen independently \citep{simpson:17}.
\item The model is highly sensitive to prior choices for the precision
    parameters, especially the input parameters for the precision of
    the spatial field \citep{illianal-b:12, beguin:12, sorbye:13,
        papoila:14, homburger:15}. These are not easily controlled
    using traditional hyperpriors, like Gamma distributions.
\item If the spatial component is intrinsic, the model gives different
    degrees of smoothing for different grid resolutions and priors
    have to be adjusted if the grid resolution is changed
    \citep{sorbye:13}.
\end{enumerate}
To address these problems, we suggest a subjective, application-driven
approach which facilitates clear interpretation of model components
and hyperprior specifications. This paper introduces the use of a
reparameterised model for log-Gaussian Cox processes, where the
log-intensity is expressed by
\begin{equation}
    \eta(s) = \beta_0+\sum_{j=1}^{n_{\beta}}\beta_j z_{j}(s) +\frac{1}{\sqrt{\tau}}\left(\sqrt{\phi} u^*(s) +\sqrt{1-\phi} v(s)\right), \quad \phi \in (0,1). \label{eq:model}
\end{equation}
Here, $\mm{u}^*=\{u^*(s),\,s\in \Omega\}$ represents a scaled spatial random field, and the
hyperparameters $\tau$ and $\phi$ are assigned penalised complexity
(PC) priors \citep{simpson:17}. We detail below why this modelling approach has important
advantages and avoids the three problems listed above.

First, the identifiability issue between the spatially structured and
unstructured fields is avoided, combining the two fields as one random
component with a marginal variance governed by a common precision
parameter $\tau$. Conditioned on $\tau$, the parameter $\phi$ helps
the interpretation, as it explains how much of the random variation is
attributed to the spatial term. The two hyperparameters $\tau$ and
$\phi$ are then seen to have orthogonal interpretation, which makes it
natural to choose the hyperpriors for these parameters independently,
see \cite{simpson:17} and \cite{riebler:16} for a similar formulation
for the BYM-model \citep{besag:91}. This makes hyperprior selection
transparent and we avoid tuning two precision parameters.

The second issue to consider is prior sensitivity. Prior choices for
hyperparameters in Bayesian hierarchical models are crucial as these
govern the variability and strength of dependence for the underlying
latent field. Especially, sensitivity to hyperprior choices for
precision parameters of random effects represents a major challenge
\citep{roos:11} and commonly used Gamma priors are not necessarily
appropriate \citep{lunn:09}. The prior choice for the precision
parameter $\tau$ in (\ref{eq:model}) will have a major influence on
posterior results and it is essential that we are able to interpret
the results in terms of prior information. This is facilitated using
the framework of PC priors
\citep{simpson:17}, in which the informativeness of hyperpriors are
adjusted in an intuitive way by user-defined scaling parameters.  \citet{klein2016scale} found 
that PC priors have better empirical performance than other standard
choices in a similar context.

Thirdly, the spatial field $\mm{u}^*$ in (\ref{eq:model}) needs to be
scaled to ensure that a given prior for its precision yields the same
degree of smoothness for different grid resolutions. This is
automatically the case for Gaussian (Markov) random field priors
\citep{cressie:93,rueheld:05}. However, if the spatial model is
intrinsic, the marginal variance will depend on the grid resolution
and the model needs to be scaled to give a unified interpretation for
its precision parameter. One approach to avoid this issue is to
scale the model in terms of its generalized variance
\citep{sorbye:13}.

\begin{figure}[t]
    \begin{center}
           \includegraphics[width=0.4\textwidth]{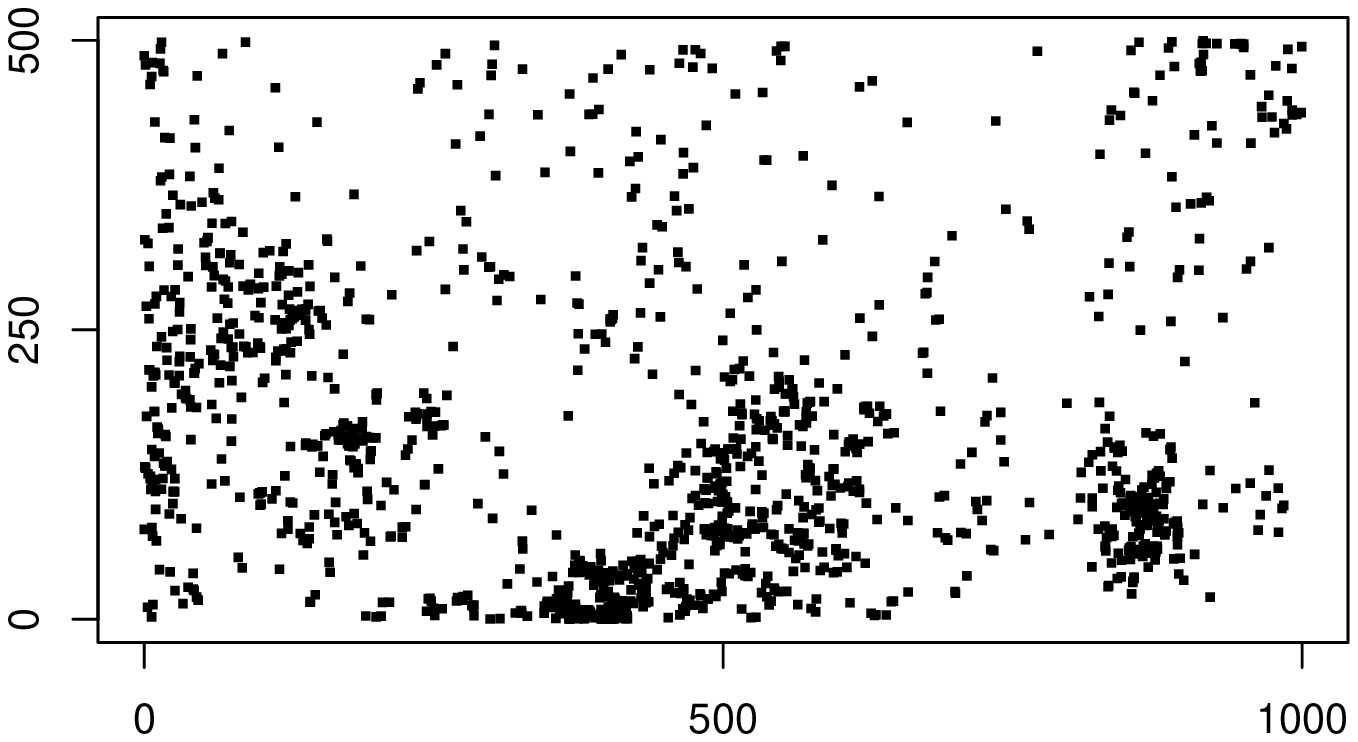}
             \includegraphics[width=0.4\textwidth]{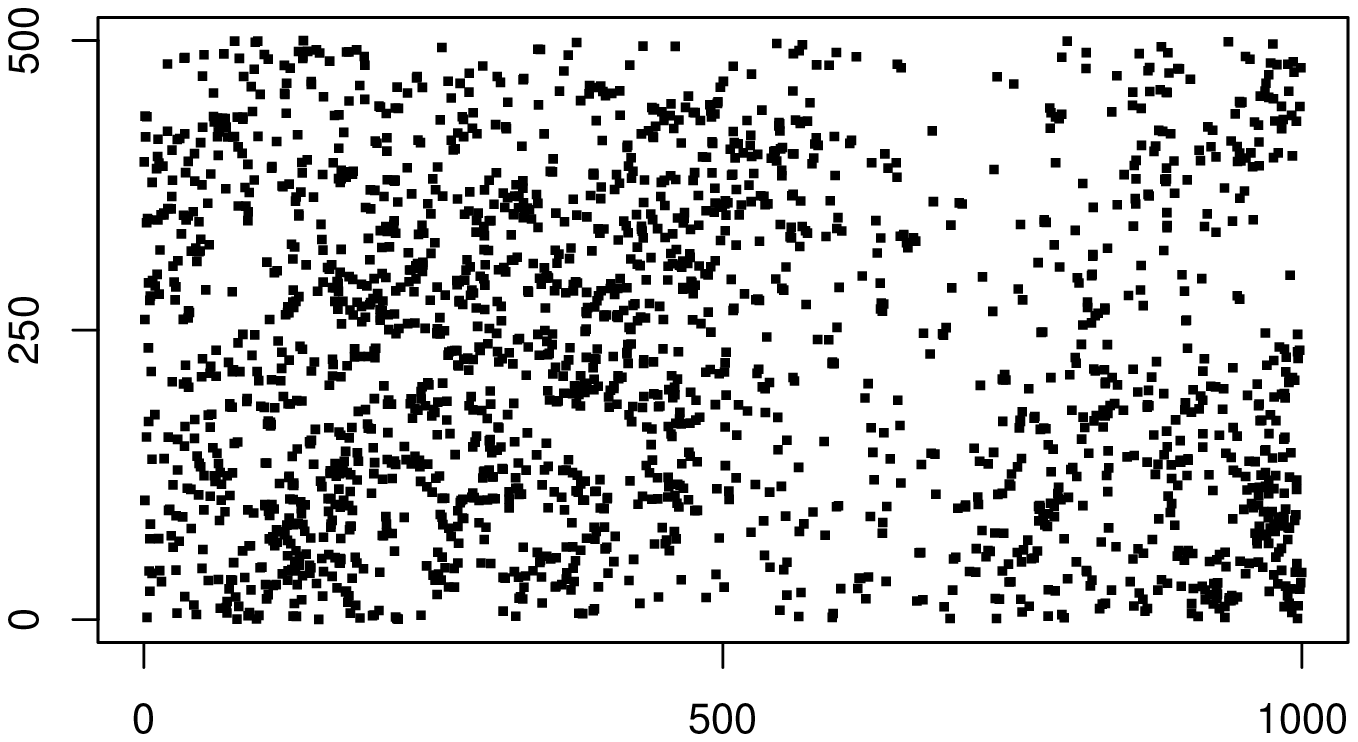}
          \caption{
            The point pattern formed by the species {\it Calophyllum Longifolium} (left) and {\it Oenocarpus mapoura} (right)  observed in a 50-ha study plot on Barro
            Colorado Island, Panama.}
            \label{fig:pattern}
    \end{center}
\end{figure}

\begin{figure}[ht]
    \begin{center}
              \includegraphics[width=0.4\textwidth]{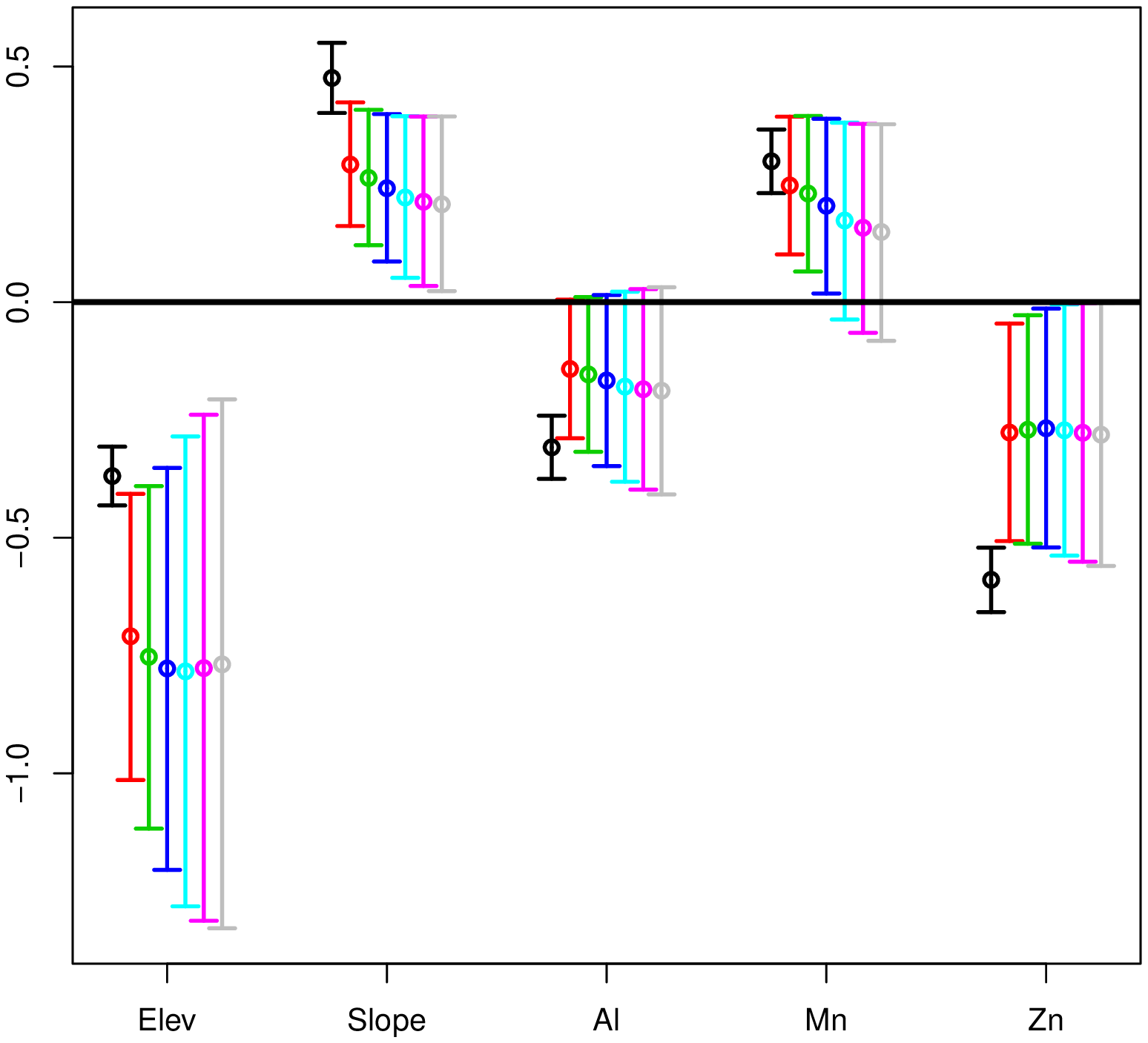} 
               \includegraphics[width=0.4\textwidth]{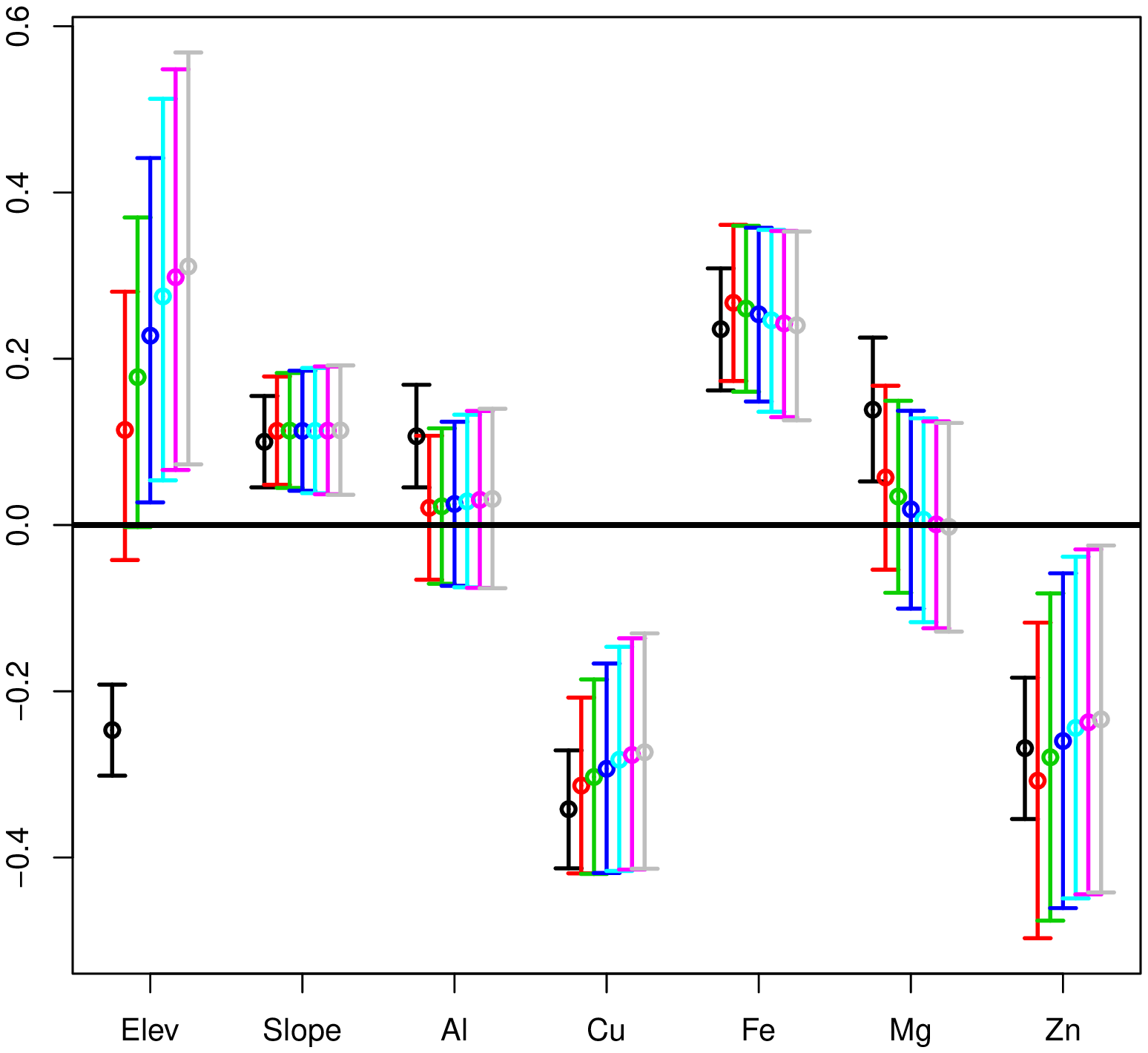} 
        \caption{The posterior mean and $95\%$
            point-wise credible intervals for the linear effects of
            fixed covariates for  \textit{Calophyllum longifolium} (left) and  \textit{Oenocarpus mapoura} (right). The black intervals show results using an ordinary generalized linear model with only linear effects of the covariates. For the other intervals, the PC prior for $\tau$ in
            (\ref{eq:model}) is scaled in terms of a user-defined
            scaling parameter $U_{\sigma}$, prescribing a maximum value for the
            marginal standard deviation $\tau^{-1/2}$, here using $U_{\sigma}=0.05$ (red), $U_{\sigma}=0.10$ (green), $U_{\sigma}=0.20$ (blue), 
            $U_{\sigma}=0.5$ (cyan), $U_{\sigma}=1.0$ (purple) and $U_{\sigma}=2.0$ (grey).}
        \label{fig:credint}
    \end{center}
\end{figure}

To initially illustrate our point, consider the point patterns formed
by the species \textit{Calophyllum longifolium} (1461 trees) and \textit{Oenocarpus mapoura} (2027 trees) (Figure~\ref{fig:pattern}), using the 2005 census of the BCI data.
Figure~\ref{fig:credint} illustrates the posterior $95\%$ point-wise
credible intervals for each of a number of selected covariates included as
fixed effects in (\ref{eq:model}). Using an ordinary generalized linear model (GLM) with only fixed effects, 
all of the covariates are significant (black intervals). However, the variance is underestimated giving too narrow credible intervals  
as spatial autocorrelation is ignored. 
When the random field component is included,  it is essential that the hyperprior
specifications can be controlled and the implication on the interpretation of the model are well-understood as these govern the statistical
conclusions made, as is clear from the figure. This is achieved using a range of hyperprior models for $\tau$, 
governed by a  scaling parameter which prescribes a maximum value
for the marginal standard deviation $\sigma=\tau^{-1/2}$ (see
Section~\ref{sec:modelling}).

Further details on the modelling approach taken here are given in
Section~\ref{sec:modelling}, including scaling, computation and tuning
of hyperpriors, and specifications for running the proposed model in
the \texttt{R}-package \texttt{R-INLA}. Section~\ref{sec:analysis}
contains a thorough discussion on fitting the proposed model to the
given example pattern, including interpretations of covariate associations 
and the posterior random fields. Discussion and
concluding remarks are given in Section~\ref{sec:discussion}. 
A brief review on the principles underlying computation of penalised
complexity priors is given in Appendix~\ref{appendix:pc}.

\section{Modelling details} \label{sec:modelling}

Log-Gaussian Cox processes may be interpreted as latent Gaussian
models \citep{rueal:09} where the observations are conditionally
independent, given a latent Gaussian field $\mm{\eta}=\{\eta(s),\,s\in \Omega\}$. Here, we consider the case where
the bounded observation window $\Omega$ is gridded and the set
$\{y_i\}_{i=1}^n$ denotes the number of points in $n$ grid cells
$\{s_i\}_{i=1}^n$. The conditional distribution of the counts is then
assumed to approximately follow a Poisson distribution, i.e.,
\begin{equation}
    y_{i}| \eta(s) \sim \mbox{Poisson}\left(\int_{s_{i}}\exp(\eta(s))ds\right)\approx \mbox{Poisson}(|s_{i}| \exp(\eta(s_i))).\label{eq:poisson}
\end{equation}
The area of a grid cell is $|s_i|$ and $\eta(s_i)$ denotes a
representative value of the point log-intensity in cell $s_i$. Since
the latent field $\eta(.)$ may be chosen flexibly, log-Gaussian Cox
processes can be extended to analyse complex spatial point patterns,
for example replicated point patterns and point patterns with marks
\citep{illianal-a:12, illianal-b:12, illianal:13}, and also to non-regular data \citep{simpsonal:16}.
However, the fitting of these models involves challenges, especially in terms of
how hyperprior assumptions influence the inference. Unless we can
communicate the meaning and interpretation of the role of priors for
the hyperparameters, we cannot appropriately communicate the role of
the latent model itself.

\subsection{Spatial field specifications including scaling}
Consider the case where the spatially structured field
$\mm{u}=(u_1,\ldots , u_n)$ in (\ref{eq:model}) is an intrinsic
conditional auto-regressive (ICAR) model \citep{besag:91}. This prior is commonly 
applied to model underlying spatial dependence structures between
neighbouring observations in Bayesian hierarchical models \citep{banerjee:04, assuncao:09}. 
The ICAR models are not proper but have several beneficial properties \citep{besag:95}. 
Especially, these models fit very nicely within the framework of PC priors 
as the models can be seen to penalise local deviation from their null space, as explained below. 

Specifically, consider the  ICAR prior defined on a regular lattice, also referred to as a second-order
intrinsic Gaussian Markov random field (IGMRF) 
\citep[Section 3.4.2]{rueheld:05}. The density of this model is defined by
\begin{equation}
    \pi(\mm{u}\mid \tau_u)\propto \tau_u^{\frac{1}{2}(n-2)}\exp{\left(-\frac{1}{2}\mm{u}^T\mm{Q}\mm{u}\right)}, \label{eq:igmrf}
\end{equation}
where the precision matrix $\mm{Q}=\tau_u \mm{R}$. The parameter $\tau_u$ denotes the precision while 
$\mm{R}$
reflects the specific neighbourhood structure of the model. The
precision matrix is singular of rank $n-2$ and does not specify an
overall value for the mean, nor a finite variance. This is not a problem, as one can impose linear
constraints to make the  variance finite and the model then specifies
a valid joint density. In fact, the rank deficiency implies that the null space of the model is a plane \citep{rueheld:05} 
and the given model penalises local deviation from this plane. The marginal variances of the model, 
integrating out the random precision $\tau_u$, give information on how large we allow this local deviation to be. 
This implies that the hyperprior for $\tau_u$ has a clear interpretation, at least for a given model of a specified dimension.  

An important feature of the IGMRF model which has usually not been accounted
for in spatial point process analysis \citep{rueal:09, illianal-a:12,
    illianal-b:12, illianal:13, kang:14}, is that the marginal
variances of the model depend on both the size and structure of
$\mm{R}$ \citep{sorbye:13}. This implies that a chosen prior for
$\tau_u$ gives different degrees of smoothing, for different grid
resolutions. To ensure a unified interpretation of the precision
parameter $\tau_u$, the model $\mm{u}$ needs to be scaled. One way to do
this is to consider the generalized variance of $\mm{u}$, computed as
the geometric mean
\begin{eqnarray*}
  \sigma^2_{\mbox{\scriptsize{GV}}}(\mm{u}) = \frac{1}{\tau_u}\exp\left(\frac{1}{n}\sum_{i=1}^n \log(\sigma^2_{ii})\right),
\end{eqnarray*}
where $\sigma^2_{ii}$ are the diagonal elements of the generalized
inverse of $\mm{R}$ \citep{sorbye:13}. The generalized variance can be
seen to represent a characteristic or typical level for the marginal
variances. For the given second-order IGMRF on a lattice, this
generalized variance will increase by a factor of $k^2$ if the
resolution for a study region is refined from $n\times n$ to
$kn\times kn$ grid cells \citep{lindgren:11}. By scaling
$\mm{R}$ to have a generalized variance equal to 1 \citep{sorbye:13}, the hyperprior for $\tau_u$ is invariant to grid
resolution. 

\subsection{Tuning hyperparameters using PC priors}
By using a  scaled IGMRF-model  $\mm{u}^*$,  
the random field component in
(\ref{eq:model}), defined by
\begin{equation}
    \psi(s)=\frac{1}{\sqrt{\tau}}\left(\sqrt{\phi} u^*(s) +\sqrt{1-\phi} v(s)\right)\label{eq:rcomp}
\end{equation}
is also scaled. This ensures that  the precision parameter $\tau$ of the reparameterised model component 
has  the same interpretation for different grid resolutions.  
Obviously, the grid resolution can still have a minor
influence on the results as grid resolution impacts on the accuracy of
the Poisson approximation in (\ref{eq:poisson}), where a very fine
resolution yields a more accurate approximation than a coarser
resolution. However, the higher the resolution the higher the required
computation time -- a trade-off that is relevant in practice. This will be discussed further for the given data example 
in Section~\ref{sec:analysis}.

The parameter $\tau$ represents the marginal precision
of the random field component, and the prescribed size of this random effect is
governed by the hyperprior specifications for $\tau$. Further, for a given
finite precision, $\phi\in(0,1)$ can be interpreted as a mixing
parameter which blends in spatial dependence in the model. As the two
hyperparameters $\tau$ and $\phi$ control different features of the
fitted random field, it is natural to assign independent hyperpriors
to these parameters. 

To facilitate control and interpretation of hyperprior choices, we use
the recently developed framework of PC priors
\citep{simpson:17}. A main idea in calculating PC priors is that a
random component, as in (\ref{eq:rcomp}), can be seen as a flexible
version of a simpler base model.  An exponential prior is then assigned to a
univariate measure of deviance from the flexible model to the simpler
model, and transformed to give the PC prior for the parameter of
interest. An important motivation for the given reparameterisation 
is to provide a proper sequence of base models, which can be used to
derive PC priors for the hyperparameters $\phi$ and $\tau$, separately. Here, the
simplest base model is one with no random effect, which corresponds to
infinite precision $\tau$. This implies that given the covariates the
pattern exhibits complete spatial randomness as represented by a
homogeneous Poisson process. For a finite value of $\tau$, an
alternative base model is defined as having no spatially structured
effect ($\phi=0$). This could for example correspond to over-dispersed
point patterns, where over-dispersion has no spatial structure and is
picked up by the error field.

\subsubsection{The PC prior for the marginal precision $\tau$}\label{sec:pc-tau}
In general, the PC prior for a precision parameter $\tau$ \citep{simpson:17} is derived by assuming no
random effect as the base model. In deriving the PC prior for $\tau$ in (\ref{eq:rcomp}), the flexible and base models are 
defined by  $(n-2)$-dimensional multinormal densities,
$f_1=\pi_G(0,\tau^{-1}\mm{\Sigma})$ and
$ f_0=\pi_G(0,\tau_0^{-1} \mm{\Sigma})$, respectively. Here,
$\tau<<\tau_0$, where $\tau_0$ is a large fixed value and the
covariance matrix is
$$\mm{\Sigma}=\phi \mm{R}^{-1}+(1-\phi)\mm{I},$$
where $\mm{R}^{-1}$ is the generalized inverse of $\mm{R}$ in
(\ref{eq:igmrf}). The deviation from $f_1$ to the base model $f_0$ is
defined by the univariate distance measure
$$d(\tau)=\sqrt{2\mbox{KLD}(f_1|| f_0)}=\sqrt{\frac{(n-2)\tau_0}{\tau}}, $$
where KLD is the Kullback-Leibler divergence \citep{kullback:51}.
Using a principle of constant rate penalisation \citep{simpson:17},
the distance $d(.)$ is assigned an exponential distribution with rate
$\theta$. Note that this automatically gives a prior on the distance,
with its mode at the base model. By a change of variables, the
resulting prior for $\tau$ is the type-2 Gumbel distribution, i.e.,
\begin{equation*}
    \pi(\tau)=\frac{\lambda}{2} \tau^{-3/2}\exp{\left(-\frac{\lambda}{\sqrt{\tau}}\right)},\label{eq:gumbel}
\end{equation*}
where $\lambda=\theta \sqrt{(n-2)\tau_0}$, and where $\theta$ is kept
constant when $\tau_0\rightarrow \infty$. 
This density corresponds to using an exponential prior on the standard deviation $\sigma=\tau^{-1/2}$.

The parameter $\lambda$ controls how fast the model shrinks towards
the base model and to govern the size of (\ref{eq:rcomp}) it is
essential that reasonable values of $\lambda$ can be inferred
intuitively. A natural suggestion \citep{simpson:17} is to impose an upper limit $U$
for the marginal standard deviation,
\begin{equation}
    P(\sigma>U_{\sigma})=\alpha_{\sigma}, \label{eq:U}
\end{equation}
where $\alpha_{\sigma}$ is a small probability. This implies that $\lambda=-\ln \alpha_{\sigma}/U_{\sigma}$ where 
$U_{\sigma}$ is a user-defined scaling parameter. To suggest reasonable values
of $U_{\sigma}$, we notice that the marginal standard deviation of a random
effect $\mm{w}\sim N(0,\tau^{-1}\mm{I})$ is about $0.31U_{\sigma}$, when $\tau$
is integrated out. Assuming that the value of $\mm{w}$ is within three
times the standard deviation, it seems reasonable that $U_{\sigma}$ represents
a prescribed maximum value for the model component $\psi$ in
(\ref{eq:rcomp}). Equivalently, $\exp(U_{\sigma})$ represents an \textit{a priori} chosen upper limit for the 
 point intensity for each cell that is explained by the
random component rather than the covariates.

\subsubsection{The PC prior for the mixing parameter $\phi$}\label{sec:pc-phi}
Assuming a finite precision parameter in (\ref{eq:rcomp}), a natural
base model corresponds to having a random effect which is just random
noise with no spatial structure, obtained when $\phi=0$. The PC prior
for $\phi$ \citep{simpson:17} is now derived by considering the
deviation between the flexible Gaussian model
$f_1=\pi_G(0,\phi\mm{R}^{-1}+(1-\phi)\mm{I})$ and the base model
$f_0=\pi_G(0,\mm{I})$. The resulting distance measure is
\begin{equation*}
    d(\phi)=\sqrt{\phi\left(\mbox{tr}(\mm{R}^{-1})-n\right)-\ln|\phi\mm{R}^{-1}+(1-\phi)\mm{I}|}, \label{eq:dphi}
\end{equation*}
which can be computed efficiently by embedding the model onto a larger
torus \citep[ch. 2]{rueheld:05}. In practice, the exponential prior
assigned to $d(\phi)$ is transformed to give a prior for
$\mbox{logit}(\phi)$, as $\phi$ is bounded. The resulting prior can
not be expressed in closed form, but computed numerically making use
of the standard change of variable transformation.

To complete the specification for the prior we again need to infer a
value for the rate parameter of the exponential distribution.
\cite{simpson:17} suggest to infer the rate parameter  based on
\begin{equation}
    P(\phi<U_{\phi})=\alpha_{\phi} \label{eq:phi},
\end{equation}
where $\alpha_{\phi} > d(U_{\phi})/d(1)$. For example, a reasonable formulation
might be $P(\phi < 0.5) = 2/3$, which gives more density mass to
values of $\phi$ smaller than 0.5 \citep{riebler:16}. A priori, we
then assume that the unstructured random effect accounts for more of
the variability than the spatially structured effect. The robustness
of this prior will be investigated in Section~\ref{sec:analysis}.

\subsection{Implementation in the \texttt{R-INLA} package}
The reparameterised model component combining the spatially structured
and unstructured effect is specified in R-INLA as the latent model
component \texttt{rw2diid}. This combines the spatially structured
effect on a grid (\texttt{rw2d}) with the spatially unstructured
random error term \texttt{iid}, as defined in (\ref{eq:model}). For a
given dataset (\texttt{data}) and fixed covariate (\texttt{cov}), the
resulting call to \texttt{inla} is

\begin{verbatim}
> formula = y ~ cov + f(seq(1:(2*n)), nrow = nrow, ncol = ncol,  
                model = "rw2diid", scale.model = T, 
                hyper = list(prec = list(prior = "pc.prec", 
                                      param = c(U.sigma, alpha.sigma)), 
                             phi = list(prior = "pc",
                                      param = c(U.phi, alpha.phi))))
> result = inla(formula, family = "poisson", data = data, E = Area)
\end{verbatim}
where \texttt{n} is the number of grid cells. By default, the \texttt{rw2diid}-model is implemented using the 
scaled IGMRF-model on a lattice, having generalized variance
equal to 1. This can also be specified using the option
\texttt{scale.model = T} in the model formulation. As already explained, this option is important in practice as
we can then use the exact same prior  for $\tau$ for different grid resolutions. 
If the model is not scaled, the prior needs to be adjusted to give the same
 degree of smoothness when the grid resolution is changed.

The penalised complexity priors for the precision and mixing parameter are specified
as the hyperprior choices ``pc.prec" and ``pc", respectively, in which
the user provides values for the scaling parameters and 
probabilities in (\ref{eq:U}) and (\ref{eq:phi}). By default, the PC prior for the precision  is implemented using
$(U_{\sigma},\alpha_{\sigma})=(1,0.01)$ in (\ref{eq:U}). For the mixing parameter in (\ref{eq:phi}), the upper limit is by default set equal to 
$U_{\phi}=0.5$ while $\alpha_{\phi}$ is set equal to a value close to the
allowed minimum value $d(U_{\phi})/d(1)$.

\section{Careful prior specification in practice}\label{sec:analysis}
A main motivation of this paper is to use the reparameterised model formulation in
(\ref{eq:model}) to facilitate interpretation of hyperpriors and
communicate how these influence the statistical analysis to practitioners. In this
section, we study how covariate associations depend on hyperprior
choices and propose some simple interpretations of the estimated
hyperparameters and model components in the context of habitat assciation models for several rainforest species. We also consider how a change in 
grid resolution would influence the results. 

\subsection{The data set and covariate selection}
We consider the spatial patterns formed by the rainforest species \textit{Calophyllum longifolium} and \textit{Oenocarpus mapoura} from the BCI-rainforest dataset, 
introduced in Section~\ref{sec:intro}; see Figure~\ref{fig:pattern}. These species were chosen since there was an expectation that habitat associations and seed dispersal patterns were different for the species, based on previous work  \citep{harms:01, muller2008interspecific}.   In addition to spatial locations of a large number of tree species, 
the  dataset includes measurements on two topographical variables, terrain elevation and slope, and thirteen different soil nutrients,
which can be included as fixed covariates.  Using a lattice-based approach,  the 50-ha study plot is initially gridded into $50\times 100$ grid cells. 
We fit a log-Gaussian Cox process to the data as detailed above, assuming that the number of trees in each grid
cell follows a Poisson distribution, as given by Equation (\ref{eq:poisson}).

When necessary, the observed covariates have been log-transformed to reduce skewness and all of the covariates have been standardised prior to the analysis. This justifies using the same prior on all coefficients $\mm{\beta}$ in (\ref{eq:model}), 
which  are assigned independent zero-mean Gaussian priors with precision $10^{-3}$.  As the covariates
are highly correlated, we choose to select a subset of the covariates in which
the variance inflation factor is less than 5 to avoid biased results due to multicollinearity. Also, we remove variables that are non-significant in an initial generalized linear model. The resulting set of covariates are slightly different for the two different species. For  \textit{Calophyllum longifolium}, the variable selection procedure suggested including two topographical variables (Elevation and Slope) and three soil nutrients (Al, Mn and Zn) in the model.  For the species \textit{Oenocarpus mapoura}, five soil nutrients  (Al, Cu, Fe, Mg and Zn) were included in addition to the two topographical variables.  \cite{reich:06} point out that the deletion of  scientifically relevant variables to avoid multicollinearity can  potentially 
result in a model  that is difficult to interpret. Also, the estimates of the coefficients could be biased but they should be more precise \citep{reich:06}. For example, in this case the procedure eliminated soil P concentration, which is known to be an important predictor of  tree distributions in Panama \citep{condit2013species}, but is also often correlated with exchangeable soil Al concentrations and pH \citep{brady2013nature}.
An alternative to the given variable selection is to use
principal components to summarise covariate effects based on all the variables,  which was the approach taken by \cite{john:07} for the analysis of tree distributions in response to soil resource availability on the Barro Colorado Island plot. These can
 easily be included as fixed effects in the proposed model, but here we have chosen to focus on  individual covariate effects as well as a linear combination of these.  

\subsection{Significance and interpretation of covariate associations}
With the aim of identifying habitat association the spatial point patterns formed by the two species are analysed to establish which covariates potentially influence the spatial distribution
of the observed points. We consider 
the soil nutrients and topographical variables and assess which variables have a
significant association with the local intensity of trees for each of the patterns in Figure~\ref{fig:pattern}. This
is not trivial as the results need to be interpreted with care keeping in mind the model assumptions. Specifically, we need to be aware
that statistical significances of covariate associations might be due
to model misspecification rather than ecologically
meaningful mechanisms. For example, this is likely to be the case 
when we use a simple GLM, where spatial autocorrelation is ignored. 

To make realistic conclusions
concerning covariate associations, it is essential to consider the
results for a reasonable range of prior input values. Using the model formulation in Equation (\ref{eq:model}), the scaling parameter
$U_{\sigma}$ in Equation (\ref{eq:U}) plays a key role in adjusting prior information. 
Recall that this parameter can be seen as a prescribed
maximum value for the effect of the random field component, 
explaining the variation of the log-intensity of a pattern. A very
small value of the scaling parameter corresponds to using an 
informative spatial prior model, here equal to a plane.
Often, such a model would not explain spatial autocorrelation sufficiently. The variance is then underestimated, 
giving too narrow credible intervals for the covariates and frequentistic coverage properties of the
model will be poor. A large value of $U_{\sigma}$ corresponds to using a
non-informative prior, allowing for large  local deviation from the plane. 
Used carelessly, this might produce an overfitted random effect
which masks relevant covariate effects due to spatial confounding. This trade-off in 
scaling the hyperprior for $\tau$ has a crucial effect on
the results. In contrast, the results are very robust to tuning
the hyperprior for $\phi$. This has been investigated implementing
the given model using $U_{\phi}=0.5$, varying the
probability $\alpha_{\phi}$ from the allowed minimum value to a value of 0.8.

Figure~\ref{fig:credint} summarises the posterior mean and $95\%$
point-wise credible intervals for the fixed covariate effects, using
six different values of $U_{\sigma}$ ranging from 0.05 to 2.0. Recall that $\exp(U_{\sigma}$) represents a prescribed upper limit 
for  the proportion of the point intensity of each cell, explained by the random component. For \textit{Calophyllum longifolium}, Al is non-significant using the minimum value of $U_{\sigma}=0.05$.  Mn
has a significant positive effect when $U_{\sigma}\leq 0.2$, but then turns non-significant. 
Elevation, Slope and Zn concentration remain significant also using a very high value of $U_{\sigma}$ which then corresponds to non-informative analysis.
The ecological interpretation of these results is not  straightforward and should be elicited by experts. 
In this species the relationships with topographic variables are supported by earlier analyses of associations to discrete habitats on the BCI plot, which also found that \textit{C. longifolium} was positively associated with slopes \citep{harms:01}. A previous analysis has also highlighted the potential importance of high Al and Mn concentrations on the BCI plot correlates of species distributions, although effects on \textit{C. longifolium} were not detected explicitly and the mechanisms underlying these patterns are unclear \citep{schreeg:10}.

The spatial distribution of \textit{Oenocarpus mapoura} is significantly associated
with Elevation, Slope, Cu, Fe and Zn. A previous analysis has detected a positive association of O. mapoura to a swamp located in one part of the BCI plot \citep{harms:01}, and the analyses presented here suggest that this may be accompanied by effects of slope and soil chemistry. Although associations of plant distributions with elements such as Cu, Fe and Zn are not commonly detected, analyses of the BCI plot data have determined a role for all these elements in niche structure \citep{john:07}, and effects on the distribution of individual species are therefore reasonable. Further research is required to understand the mechanisms that drive these patterns of plant distribution in response to elements that are not required in large quantities for plant growth, or may even be toxic \citep{john:07, schreeg:10}. We notice that the effect of Elevation is positive when the spatial effect is included and becomes more significant with increasing values of $U_{\sigma}$. This is probably due to confounding between Elevation and the spatial effect, thus the significance of Elevation should be interpreted with care. 

\subsection{Interpretation of estimated hyperparameters and model components}
A possible interpretation of  $U_{\sigma}$ is to view the size of this parameter as a prior measure of
misspecification in modelling the log-intensity in a given grid cell. This reflects that that inference is made we have inference in dependence on how certain we are to have included all potential causes of aggregation in the model.  The statements and conclusions we end up with 
should then be interpreted conditional on the size of the unexplained effect.
The model misspecification has two sources corresponding to the role played
by the spatially structured effect $\mm{u}$ and the random error term
$\mm{v}$ in (\ref{eq:model}). The spatial field $\mm{u}$ models
over-dispersion in cell counts due to between-cell random effects
which are not captured by other terms in the models. In essence, these
represent non-local dependencies introduced by failing to model the
mean correctly. The error field $\mm{v}$ captures the within-cell
over-dispersion, which can be due to un-modelled random effects that
might have led to increased local clustering in the point pattern. We can
then interpret the mixing parameter $\phi$ as the balance between within- and between-cell misspecification 
and hence the relative importance of local and larger scale clustering that cannot be explained by the covariates alone.

\begin{table}
\begin{center}
    \begin{tabular}{|c|lll|lll|}\hline
      & \multicolumn{3}{c|}{$50\times 100$}   & \multicolumn{3}{c|}{$100\times 200$} \\
      $U_{\sigma}$ & $\hat{\sigma}$ $\quad$  $(95\%$ CI) & $\hat{\phi}$  $\quad$  $(95\%$ CI) & DIC  &  
      $\hat{\sigma}$ $\quad$  $(95\%$ CI)& $\hat{\phi}$ $\quad$  $(95\%$ CI)& DIC \\\hline
            0.05 &  1.54 (1.39, 1.69) & 0.71 (0.63, 0.78)  & 5113  
      	      &  1.78 (1.62, 1.97) & 0.72 (0.65, 0.79) & 8153  \\  
      0.10 & 2.02 (1.79, 2.27)  & 0.83 (0.76, 0.88)  & 5066     
      	     & 2.33 (2.09, 2.59) & 0.83 (0.77, 0.87) & 8077\\
      0.20 &  2.64 (2.29, 3.03) & 0.90 (0.86, 0.94)  & 5035   
      	     & 3.07 (2.69, 3.49) & 0.90 (0.86, 0.93) & 8031\\ 
      0.50 &  3.50 (2.99, 4.10) & 0.95 (0.92, 0.97)& 5010  
                 & 4.16 (3.51, 4.86) & 0.95 (0.92, 0.97) & 7996 \\
      1.00 & 4.05 (3.42, 4.76)  & 0.96  (0.94, 0.98) & 5001   
                  &4.85 (4.08, 5.84) & 0.96 (0.94, 0.98) & 7982 \\
      2.00 & 4.41 (3.71, 5.21)  & 0.97 (0.95, 0.98)  & 4996  
                 &5.41 (4.49, 6.43) & 0.97 (0.95, 0.98) & 7975  \\\hline 
\end{tabular}
\caption{\textit{Calophyllum longifolium:} Estimated hyperparameters including $95\%$ credible intervals 
and  the deviance information criterion (DIC) for different choices of the 
scaling parameter $U_{\sigma}$, using two different grid resolutions.}
\label{tab:hyper1}
\end{center}
\end{table}

\begin{table}
\begin{center}
    \begin{tabular}{|c|lll|lll|}\hline
      & \multicolumn{3}{c|}{$50\times 100$}   & \multicolumn{3}{c|}{$100\times 200$} \\
      $U_{\sigma}$ & $\hat{\sigma}$ $\quad$  $(95\%$ CI) & $\hat{\phi}$  $\quad$  $(95\%$ CI) & DIC  & 
      $\hat{\sigma}$ $\quad$  $(95\%$ CI)& $\hat{\phi}$ $\quad$  $(95\%$ CI)& DIC \\\hline
      0.05 &  0.59 (0.49, 0.70) & 0.38 (0.23, 0.55)  & 7971  & 
      		  0.29 (0.18, 0.42) & 0.92 (0.65, 1.00) & 13137  \\ 
      0.10 & 0.74 (0.64, 0.84)  & 0.46 (0.30, 0.62)  & 7922  &  
      		 0.84 (0.71, 0.97) & 0.43 (0.28, 0.59) & 12919 \\
      0.20 &  0.85 (0.74, 0.97) & 0.55 (0.39, 0.70)  & 7902  &   
      	 0.97 (0.84, 1.10) & 0.50 (0.34, 0.66) & 12872 \\ 
      0.50 &  0.98 (0.80, 1.17) & 0.64 (0.46, 0.79) & 7890  &  
                   1.11 (0.91, 1.33) & 0.59 (0.42, 0.75) & 12846 \\ 
      1.00 & 1.04 (0.84, 1.26) & 0.68 (0.50, 0.82)  & 7886 
                 & 1.19 (0.96, 1.47)& 0.64 (0.46, 0.79) & 12839 \\
      2.00 & 1.08 (0.87, 1.33)  & 0.70 (0.52, 0.84)  & 7884  &
                 1.22 (1.03, 1.43) & 0.65 (0.48, 0.80) & 12835  \\\hline
    \end{tabular}
      \caption{\textit{Oenocarpus mapoura:} Estimated hyperparameters including $95\%$ credible intervals and  the deviance information criterion (DIC) for different choices of the scaling parameter $U_{\sigma}$, using two different grid resolutions.}
 \label{tab:hyper2}
\end{center}
  \end{table}

Table~\ref{tab:hyper1}  and \ref{tab:hyper2} summarise the estimated hyperparameters for the two species, using both a $50\times 100$ and a $100\times 200$ grid.  For illustration, we also include the
deviance information criterion (DIC) \citep{spiegelhalter:02}, which is a commonly applied criterion in Bayesian model selection.  
Obviously, this criterion can not be used to select $U_{\sigma}$ automatically as it just decreases as 
the prescribed variability of the model increases. The results illustrate that the estimated values of  $\sigma$ and  $\phi$ typically 
increase  as a function of $U_{\sigma}$.  
This is natural and suggests that as the model marginal
variance increases, the within-cell variation becomes less important
and more of the variability is explained by the spatially structured
term. It is difficult to specify a recommended balance between these two types of misspecification, but a value of $\phi$ close to 1 might indicate
that the spatially structured field dominates too much.

For the species \textit{Calophyllum longifolium}  
$\phi$ gets close to 1 rather quickly as $U_{\sigma}$ is  increasing. This may be indicating the relative unimportance of the error field  and hence the relative irrelevance of local clustering. Conversely, for the species \textit{Oenocarpus mapoura} increasing values of $U_{\sigma}$ result in medium sized $\phi$. Hence, in contrast to the what we saw for the first species,   the error field representing  local clustering plays a relatively bigger role in explaining the spatial pattern formed by the trees.

Associations between underlying biological processes and the point
pattern formed by a species are typically much more complex than what
we are able to capture with a simple statistical model. The specific
point pattern formed can for example be influenced by interaction with
other species, dispersal limitations and a mix of abiotic and biotic
processes. To gain further insight, we consider the estimated
sum of the fixed covariate effects and the posterior random fields to
see whether these enable ecologically meaningful interpretations.

\begin{figure}[h]
    \begin{center}
         \includegraphics[width=3.5cm]{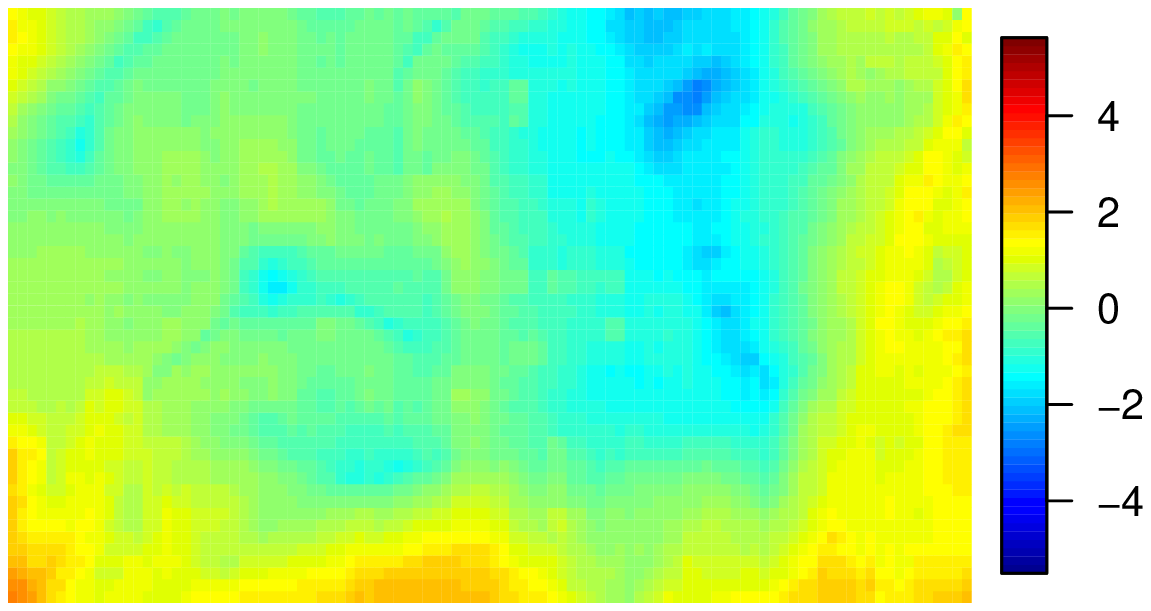}
         \includegraphics[width=3.5cm]{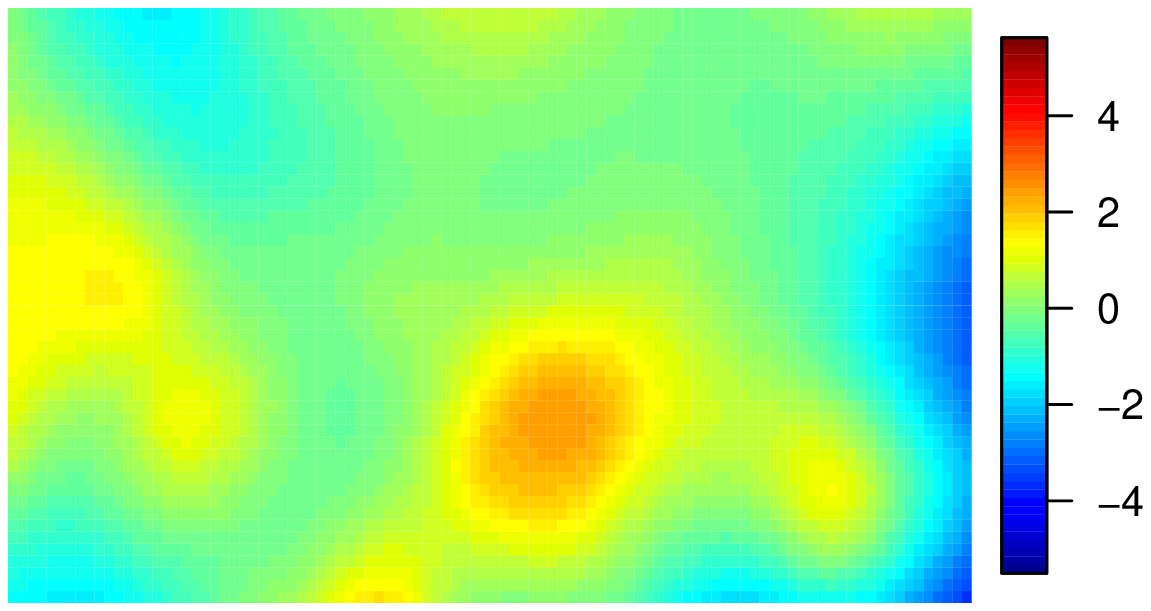}
         \includegraphics[width=3.5cm]{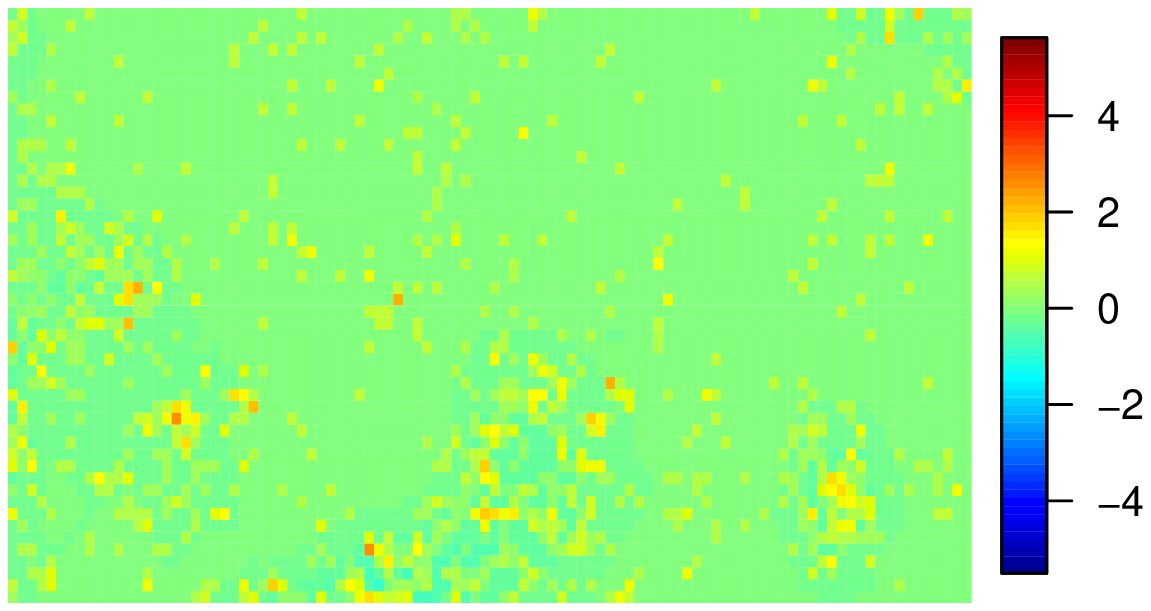}
         \includegraphics[width=3.5cm]{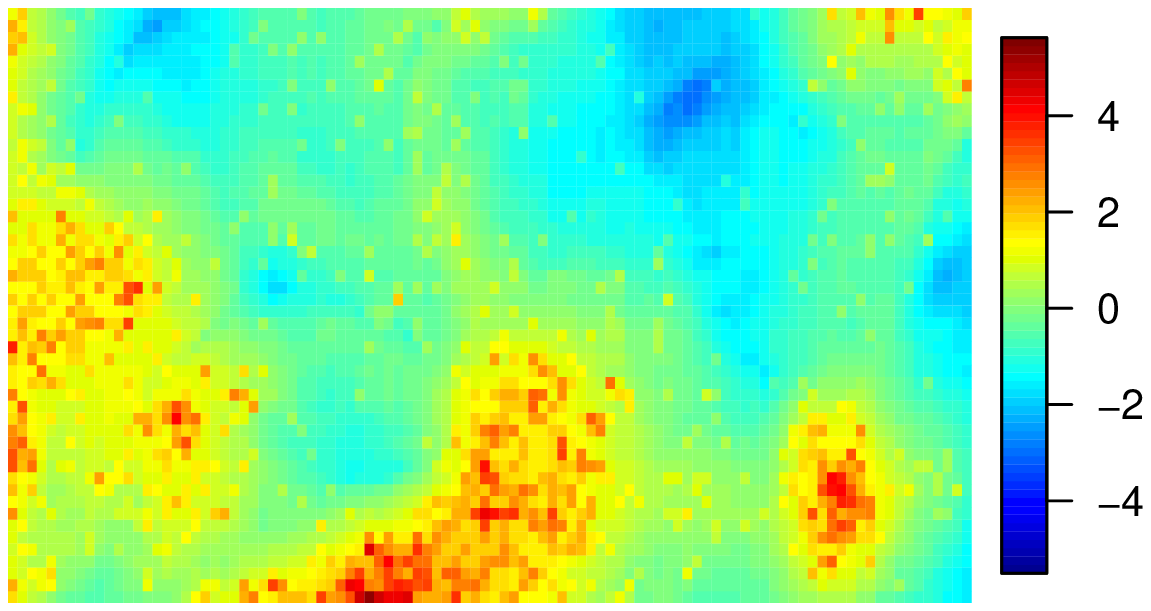}
        \\\vspace{0.3cm}
          \includegraphics[width=3.5cm]{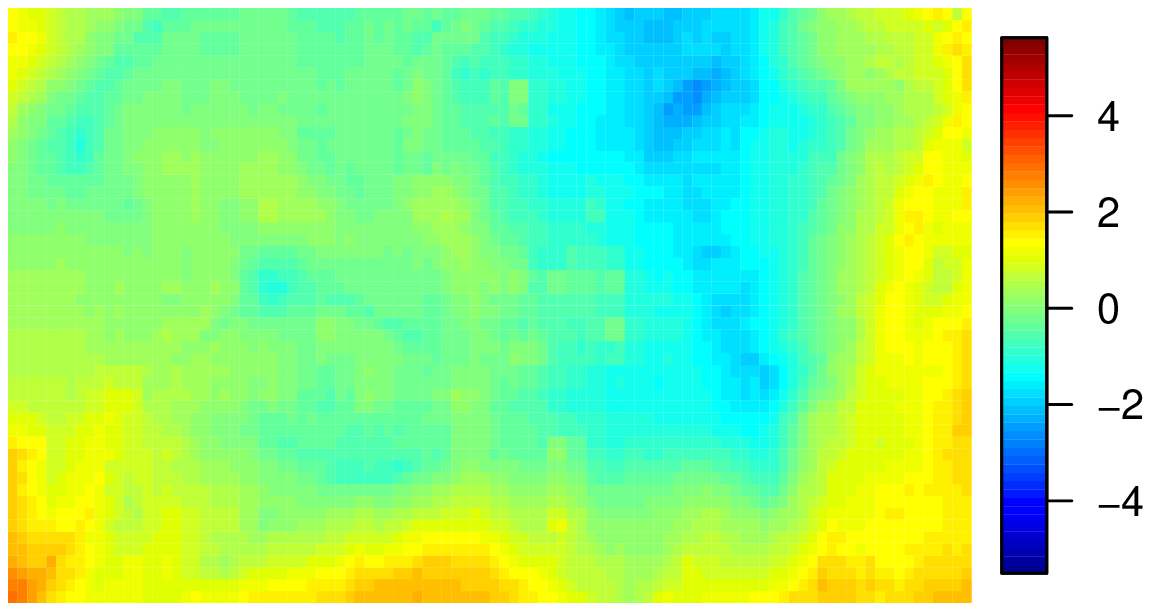}
         \includegraphics[width=3.5cm]{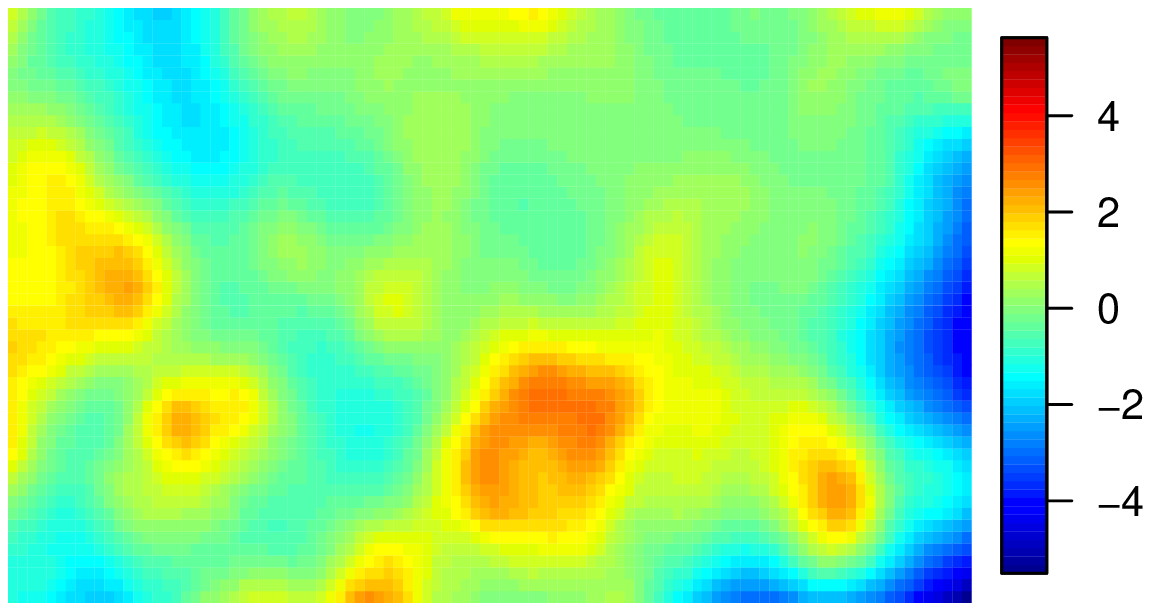}
         \includegraphics[width=3.5cm]{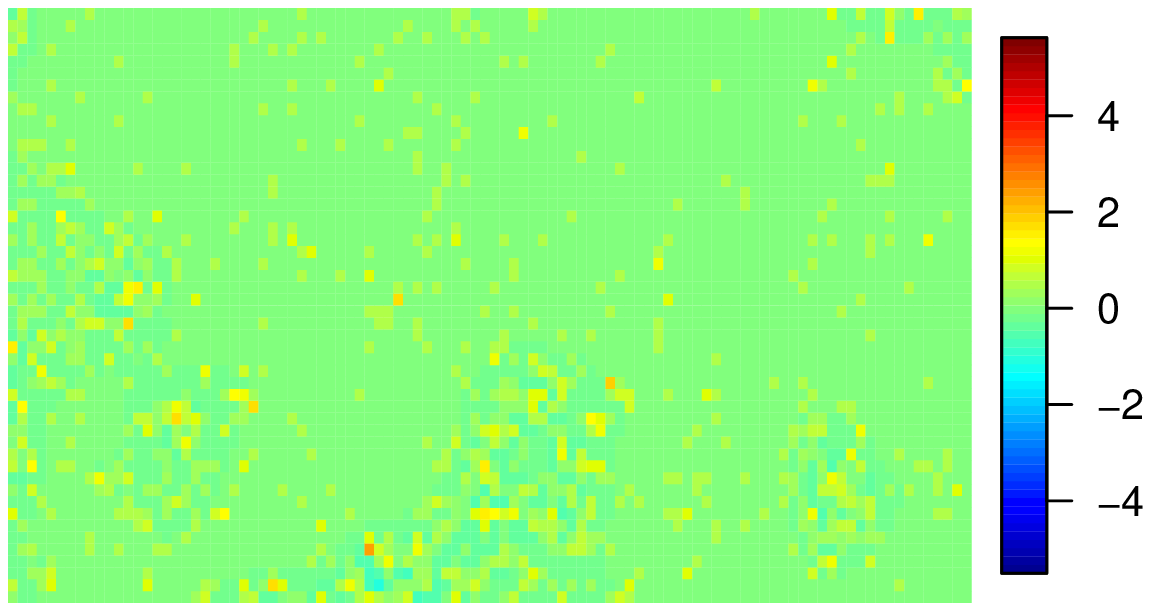}
         \includegraphics[width=3.5cm]{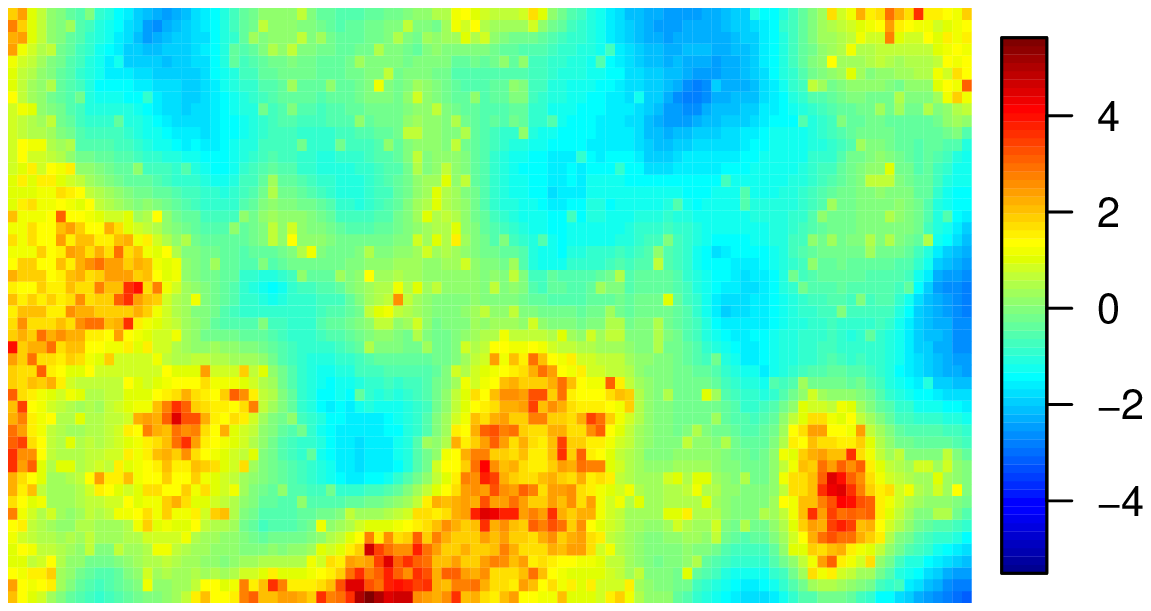}
                 \caption{\textit{Calophyllum longifolium}. The fixed covariate effects (without
            intercept), the weighted spatial component, the weighted
            error component, and the linear predictor (without
            intercept), using $U_{\sigma}=0.05$ (upper panels) and $U_{\sigma}=1.0$
            (lower panels).}
        \label{fig:calolo-effects}
    \end{center}
\end{figure}

\begin{figure}[h]
    \begin{center}
         \includegraphics[width=3.5cm]{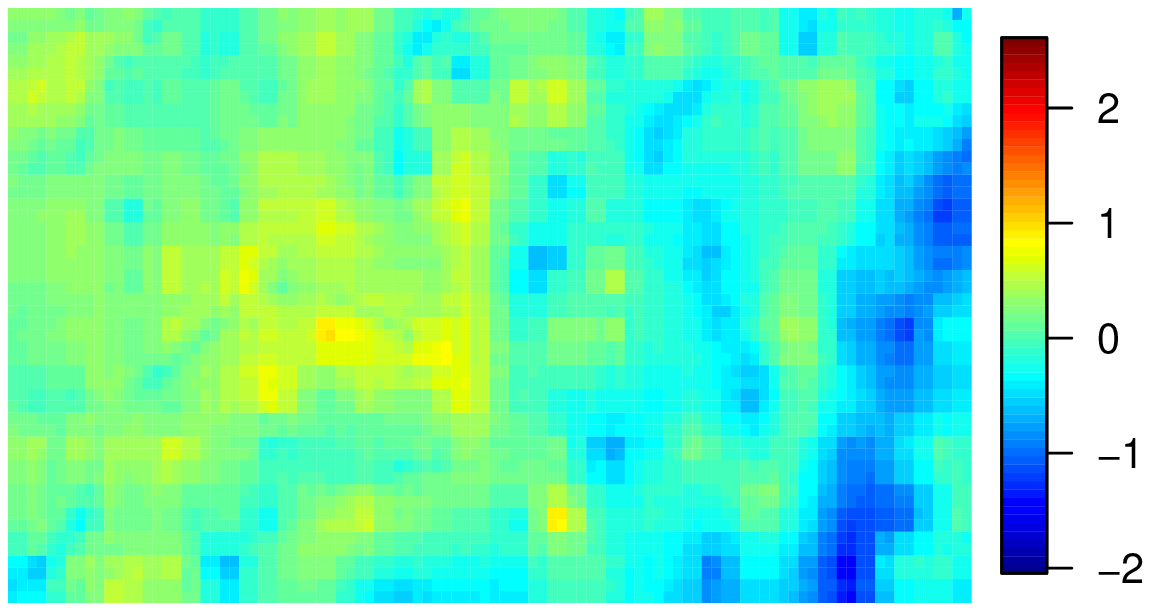}
         \includegraphics[width=3.5cm]{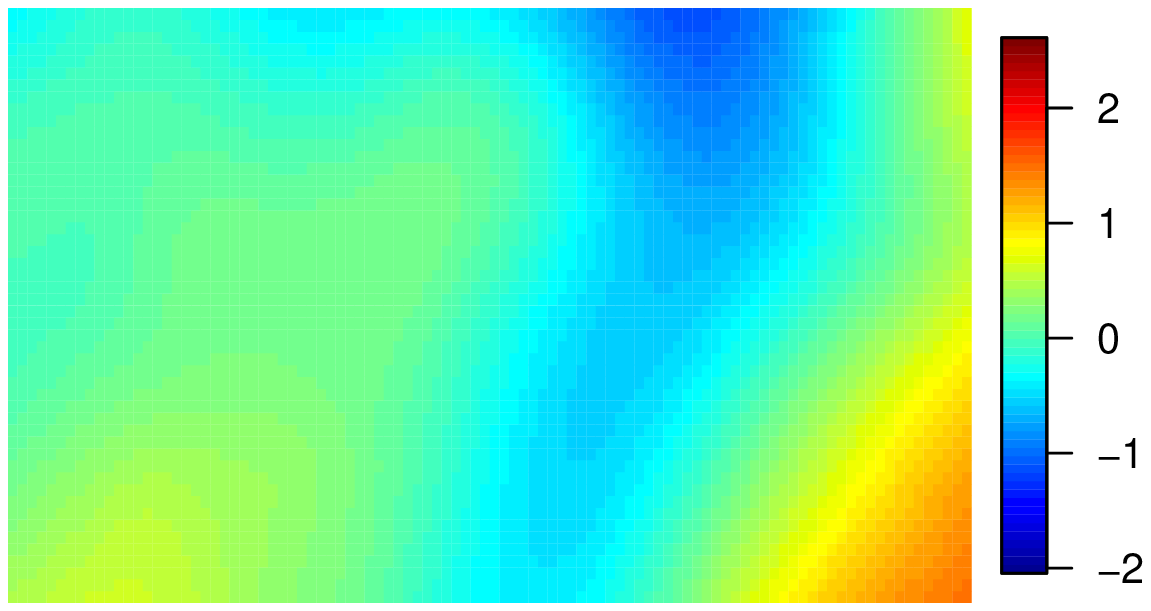}
         \includegraphics[width=3.5cm]{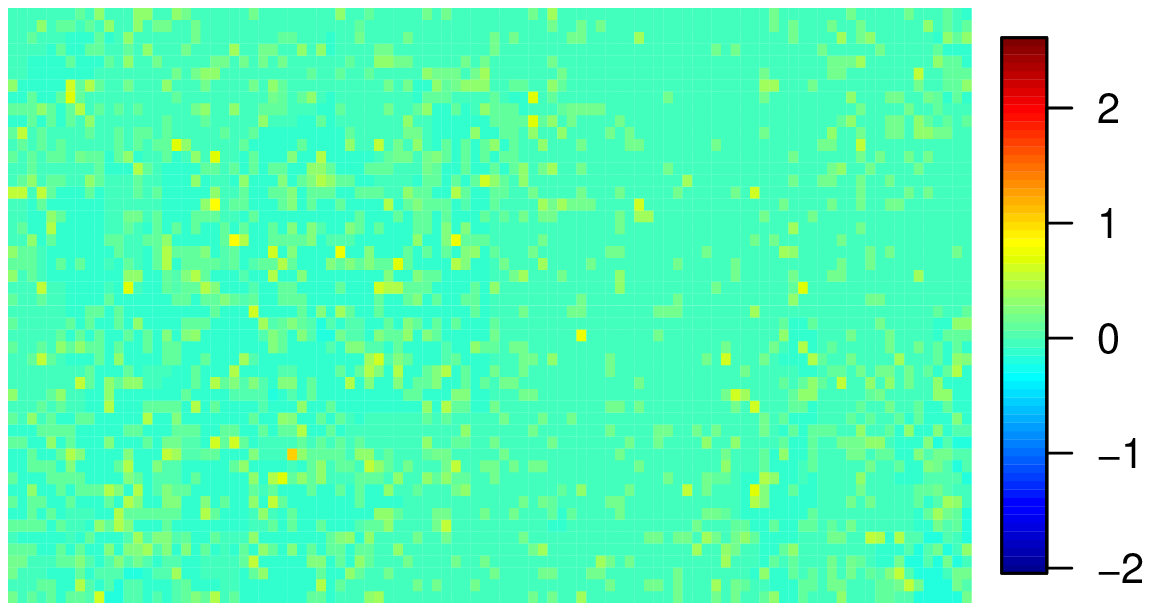}
         \includegraphics[width=3.5cm]{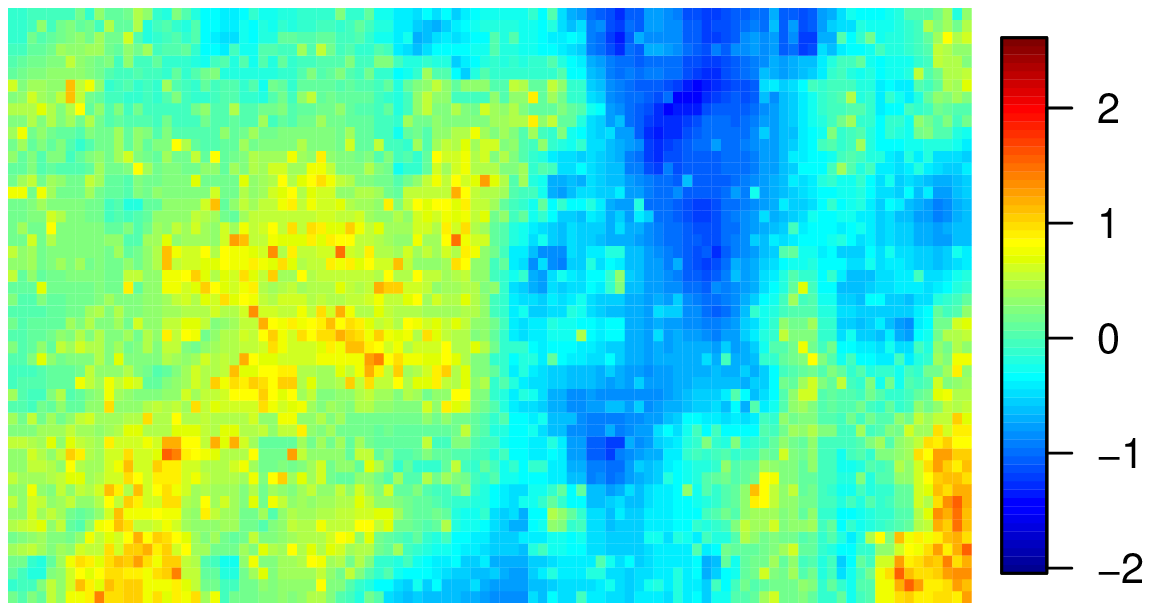}
        \\\vspace{0.3cm}
          \includegraphics[width=3.5cm]{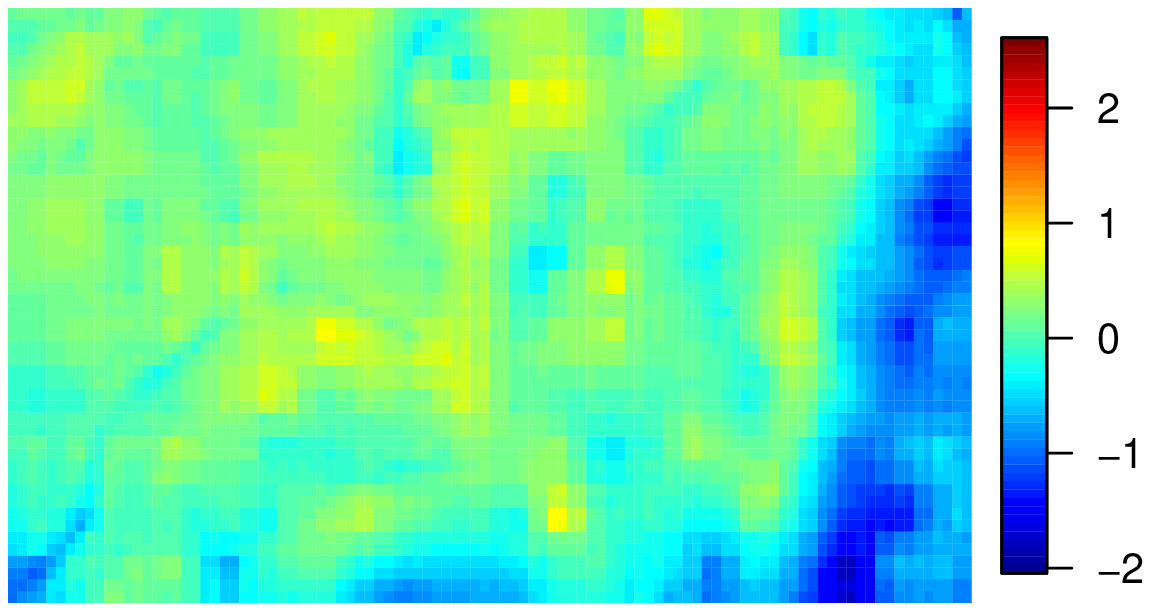}
         \includegraphics[width=3.5cm]{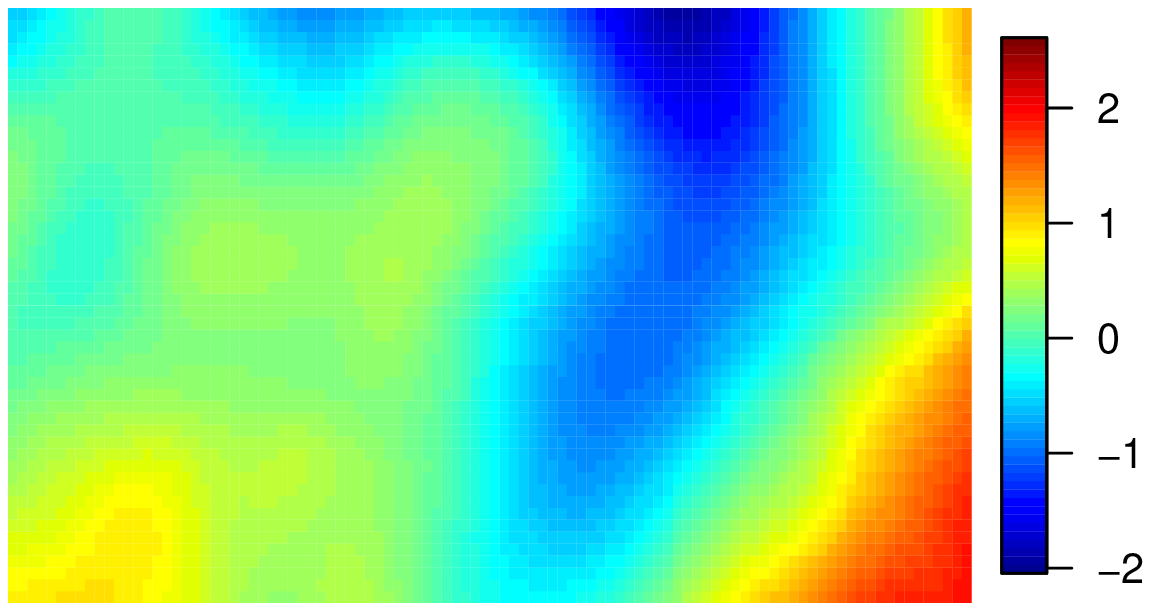}
         \includegraphics[width=3.5cm]{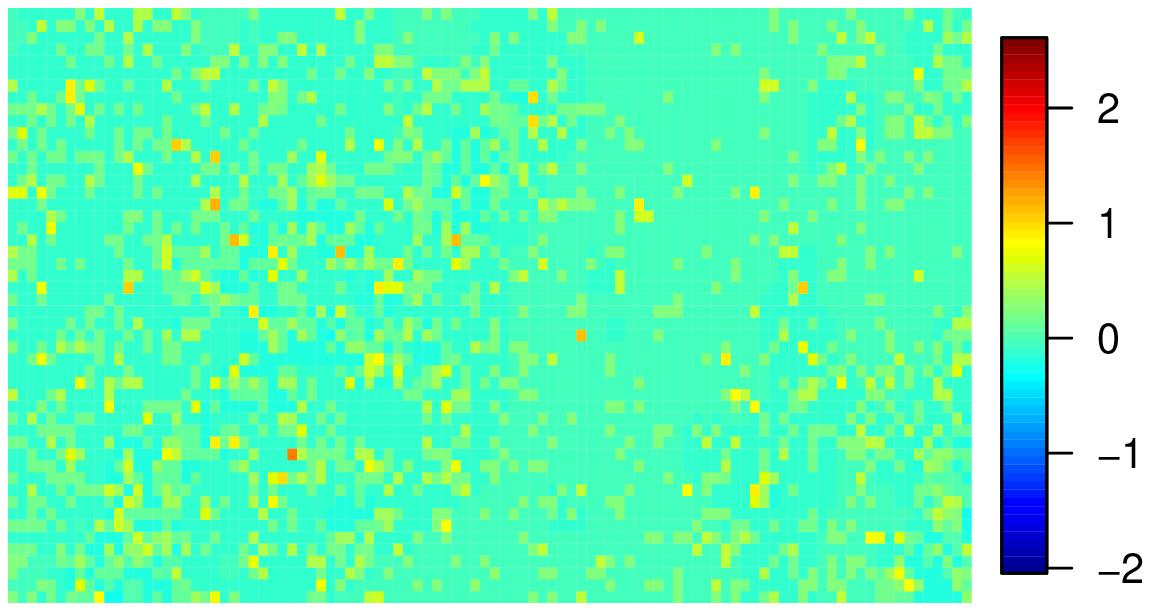}
         \includegraphics[width=3.5cm]{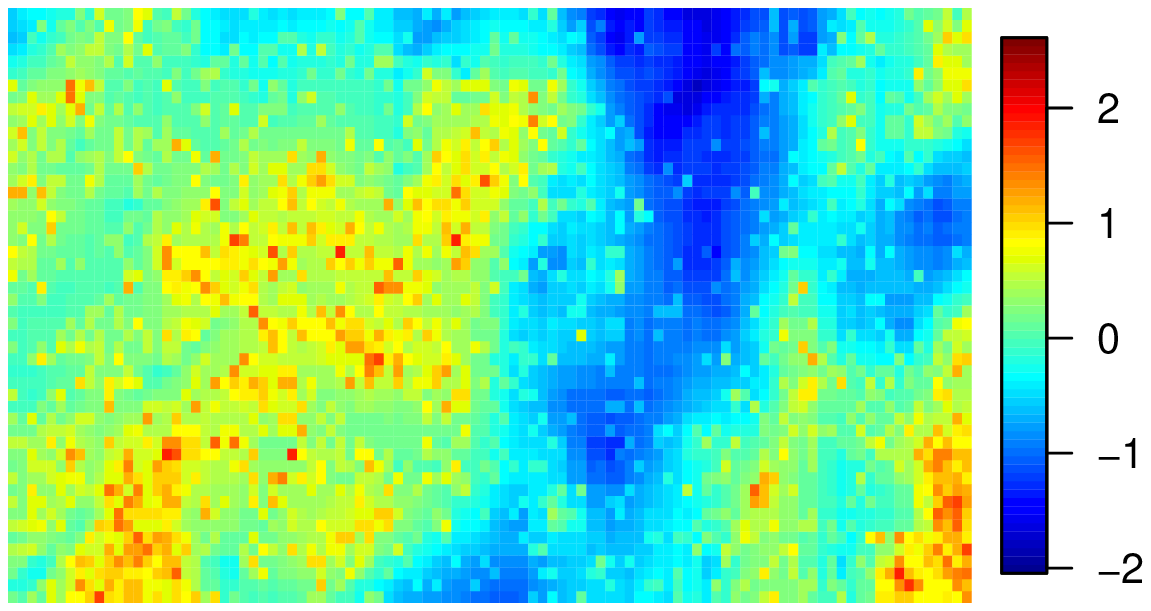}
                 \caption{\textit{Oenocarpus mapoura}: The fixed covariate effects (without
            intercept), the weighted spatial component, the weighted
            error component, and the linear predictor (without
            intercept), using $U_{\sigma}=0.05$ (upper panels) and $U_{\sigma}=1.0$
            (lower panels).}
        \label{fig:oenoma-effects}
    \end{center}
\end{figure}

Figures~\ref{fig:calolo-effects} and ~\ref{fig:oenoma-effects}  illustrate the decomposition of the
fitted log-intensity into three components -- the fixed effects of covariates, the
weighted spatially structured effect $\mm{u^*}\sqrt{\phi/\tau}$, the
weighted error term $\mm{v}\sqrt{(1-\phi)/\tau}$ and the combination of these, the resulting linear predictor. Again, it is useful to consider the results for different scaling parameters, 
here  $U_{\sigma}=0.05$ (upper panels) and $U_{\sigma}=1.0$
(lower panels).  For both species, the summarised effect of the fixed
covariates do not seem to change substantially with the different choices of the scaling parameters,  while the spatially structured
field clearly becomes more detailed. The error fields are also quite similar for both choices of $U_{\sigma}$.

\begin{figure}[h]
    \begin{center}
    \includegraphics[width=0.4\textwidth]{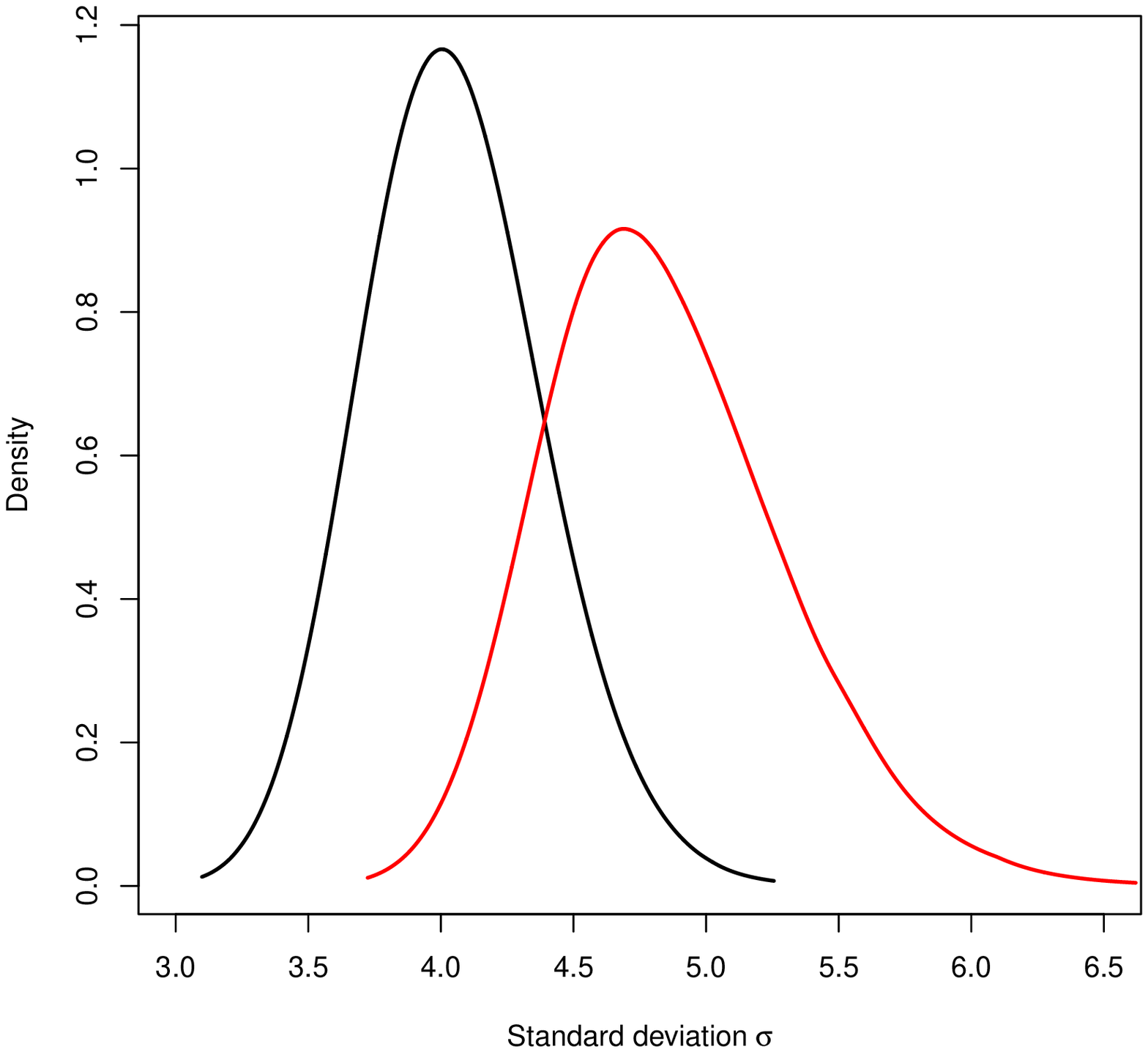}
     \includegraphics[width=0.4\textwidth]{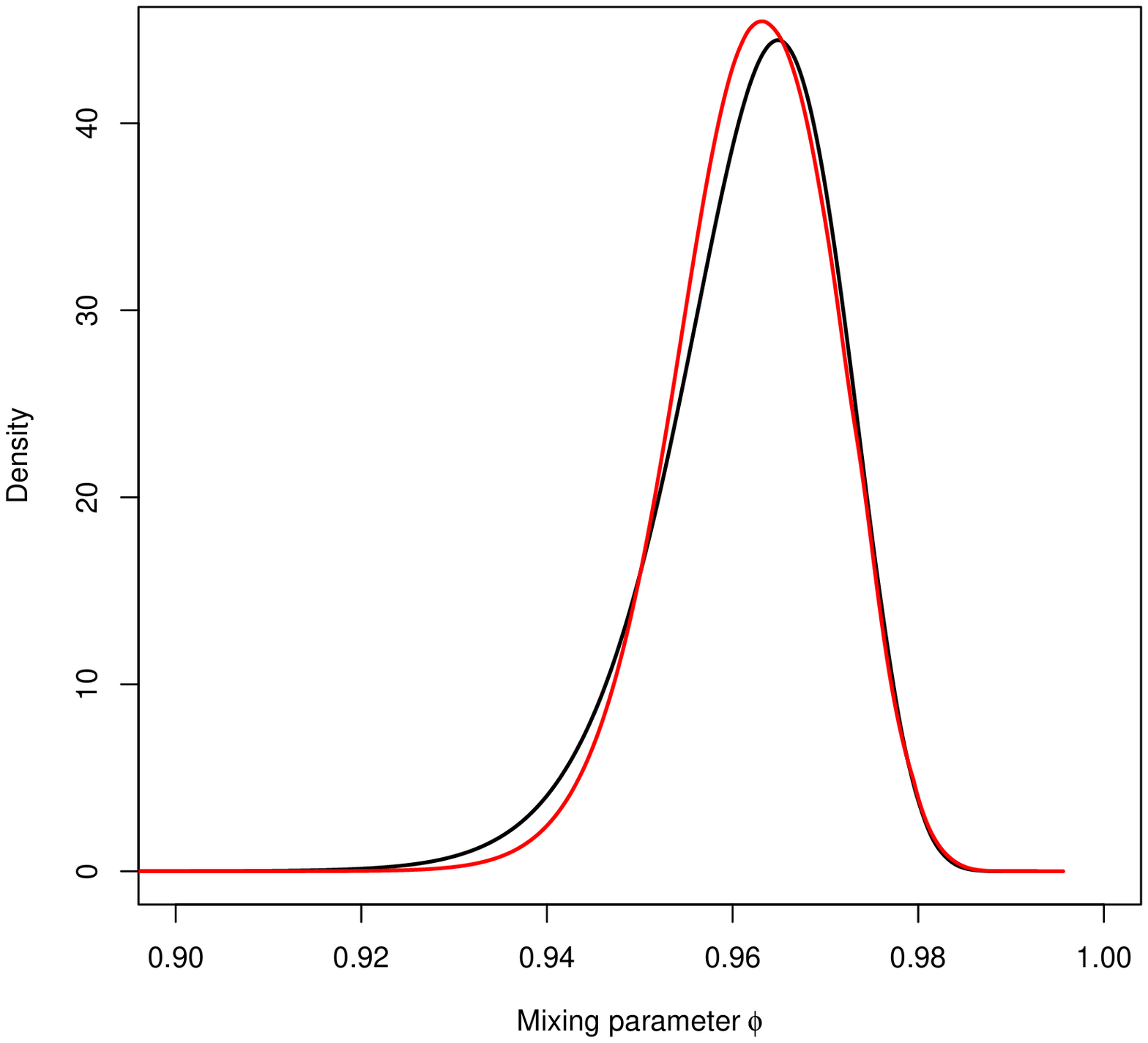}
      \includegraphics[width=0.4\textwidth]{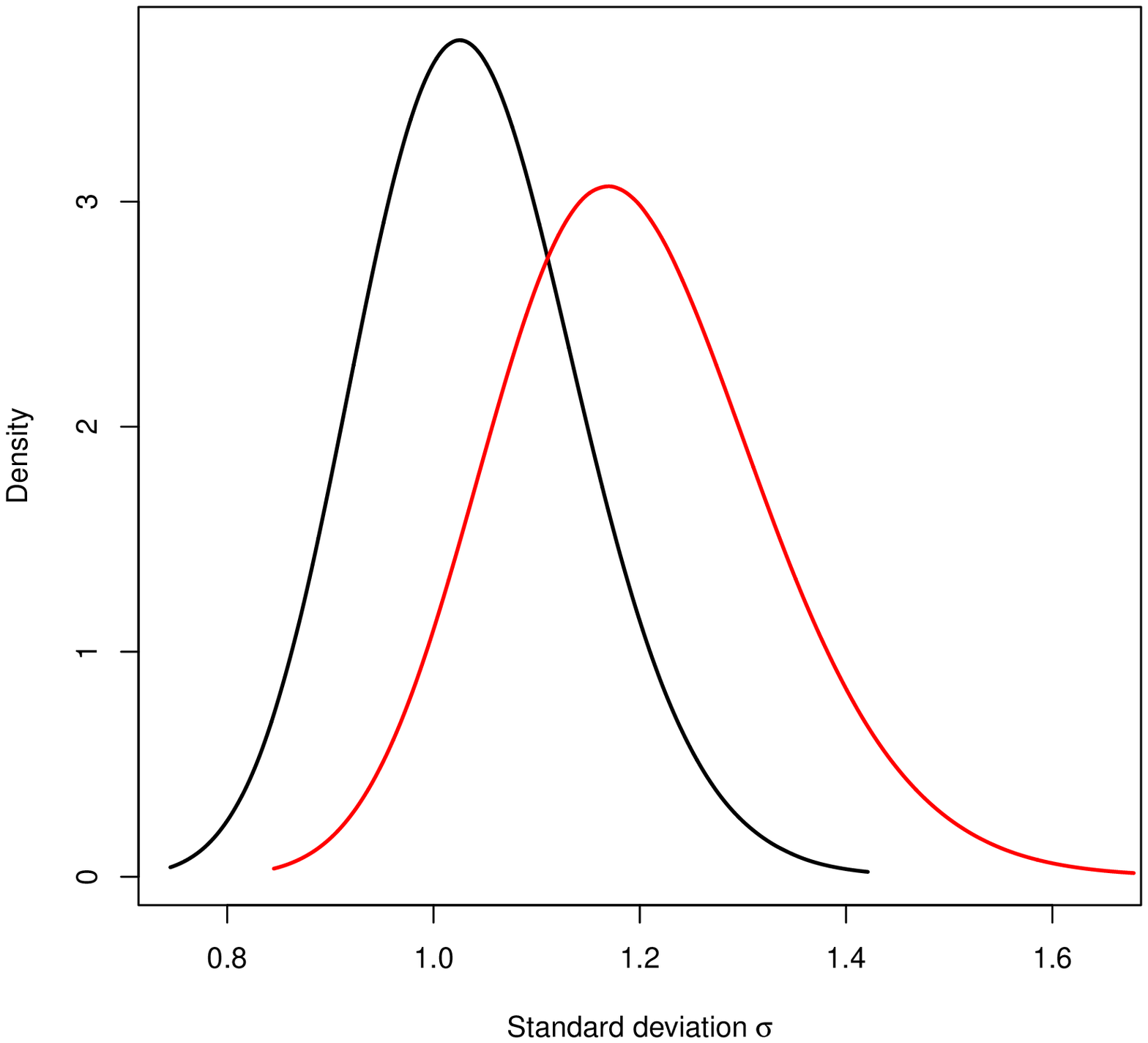}
       \includegraphics[width=0.4\textwidth]{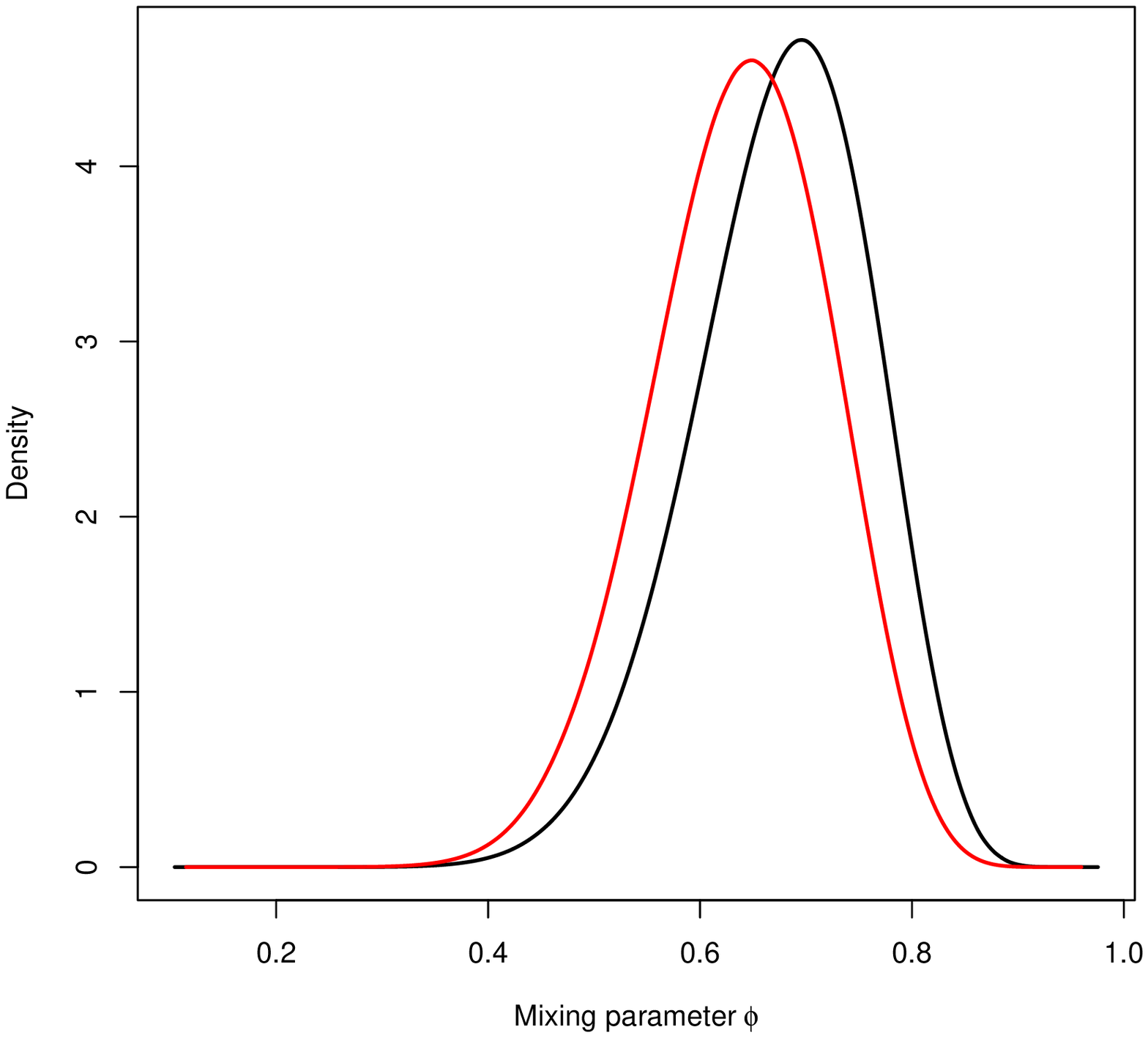}
             \caption{Posterior marginals for the standard deviation and the mixing
            parameter using grid resolution equal to $50\times 100$
            and $100\times 200$ (red), when $U_{\sigma}=1.0$ for \textit{Calophyllum longifolium} (upper panels) and \textit{Oenocarpus mapoura} (lower panels).}
        \label{fig:maquira-resolution}
    \end{center}
\end{figure}

The given results are robust to changes in the grid resolution of the study plot. This has been investigated dividing the given area into   $50\times 100$ and $100\times 200$ grid cells, respectively. The difference in running time increasing from an average of 5 minutes to almost 2 hours when moving from the rougher to the finer resolution.
By using a scaled spatial component in (\ref{eq:model}), we have ensured
that both the results summarising the significance of covariates and the interpretations of the random field
effects are similar when the  grid cell resolution is changed. The PC prior
for $\phi$ is by construction not invariant to grid resolution. However, Tables~\ref{tab:hyper1} and  ~\ref{tab:hyper2}
illustrate that the estimates of $\phi$ are very similar using the two different grid resolutions. This can also be seen in Figure~\ref{fig:maquira-resolution}, which gives the posterior marginals of both $\sigma$ and $\phi$ when $U_{\sigma}=1$, for both resolutions. 

\subsection{A note on restricted spatial regression}
A suggested alternative to the proposed method is to apply a
restricted spatial regression approach, in which the spatial field is
constrained to be orthogonal to the space spanned by the fixed
covariates \citep{reich:06, hodges:10, hughes:13, hanks:15}. This is
motivated by the fact that inclusion of spatially-correlated field in
a model with fixed effects causes shrinkage and possibly biased
estimates for the mean and variance of the fixed-effect coefficients
\citep{reich:06, hodges:10}.

We can easily implement the idea of restricted spatial regression
using (\ref{eq:model}). However, the implied assumption of no spatial
confounding between the fixed and random effects is very strong as the
observed fixed effects are attributed as much variability as possible.
This typically leads to credible intervals that are inappropriately
narrow under model misspecification \citep{hanks:15}. We also notice that the obtained
posterior means for the covariate effects are not extremely biased compared with the results obtained by GLM, 
except for the effect of Elevation for {\textit{Oenocarpus mapoura}, where it changes sign  (Figure~\ref{fig:credint}).
Finally, using a restricted spatial regression approach we loose flexibility as the effects and credible intervals for the observed
covariates are invariant to the choice of $U_{\sigma}$. 

We can not recommend this
approach, as it seems unrealistic to assume that unobserved biological
processes reflected by the random fields are completely independent of
the observed covariate information. In addition to false significances
of the fixed covariates, interpretation of the resulting random fields
is difficult. Meaningful interpretation of the posterior random
fields is important as these can be viewed to reflect underlying
biological processes and should not necessarily be considered as 
nuisance that needs to be accounted for \citep{hamel:12, illian2017improving}. They may also be used for knowledge elicitation, pointing to biological mechanisms that cause aggregation but have not been included in the model, as we will see below.

\section{Discussion and concluding remarks}\label{sec:discussion}

For the species \textit{Calophyllum longifolium}  
habitat association was expected based on its positive association with the slope habitat and negative association with the flat habitat in previous analyses of the same plot data  \citep{harms:01}. The species  is known to be well dispersed, hence local clustering was not expected and the results have confirmed this \citep{harms:01, muller2008interspecific}.  
In addition, inspection of the 
spatial field in Figure         \ref{fig:calolo-effects}
 shows clear medium scale clustering that the covariates we examined cannot explain and may be related to the activity of the diverse assemblage of vertebrate secondary dispersers and seed predators that are known to be attracted to the fruits of this species (e.g.\ \cite{adler1998fates}). Secondary dispersal and larder-hoarding of large tropical tree seeds may cause aggregations at medium scales that decouples the spatial structuring of recruitment from that of adults, which may be exacerbated by density-dependent mortality close to adult trees. These processes highlight the complexity of the interactions that determine spatial structure in tropical tree populations, including effects of biotic components of the community that are not captured by the covariates included in previous analyses. Through this example we have demonstrated that the modelling approach allows us to account for small and intermediate scale clustering while assessing the relative importance of either of these and also while  providing potential interpretation of reason for larger scale clustering unexplained by the covariates.

For  \textit{Oenocarpus mapoura} significant habitat associations were expected \citep{harms:01, muller2008interspecific} because this palm is a known specialist of the swamp habitat on the plot.
The spatial field in Figure \ref{fig:oenoma-effects}
shows a trend that covariates cannot explain, in particular, the high intensity of points in the bottom right hand corner of the plot. The cause of this cluster remains uncertain, but this corner has the lowest elevation across the plot and there may be aspects of the hydrological regime there that are particularly suitable for this species. In contrast to the other species,  local clustering was to be expected here due to known seed dispersal limitation.
This example allowed us to illustrate that we are able to establish the significance of covariate effects and their sensitivity to the smoothness (and hence potential overfitting) of the spatial field while accounting for both local and larger scale clustering and establishing their relative importance to the pattern.

An observed spatial point pattern represents one realisation of a
random spatial point process. The fact that we have only one
realisation of a random process is challenging as a statistical model
including a spatially structured random field is easily overfitted to
the observed pattern. This implies that valuable ecological
information might not be revealed due to spatial confounding. However,
underestimation of spatial autocorrelation could also give misleading
results in terms of false significances of associations between the
point pattern intensity of a species and covariates reflecting
environmental conditions. For the given log-Gaussian Cox model, this
trade-off is governed by the hyperprior choices for the precision of the 
included random field component. To avoid over-interpretation
of the results, it is essential to fit models using a reasonable range
of hyperprior choices.

In order to define reasonable hyperprior choices, the hyperparameters 
need to have clear interpretations. This is achieved here
by viewing the spatially structured and unstructured random fields as
one random component, parameterised in terms of two hyperparameters.
The common precision parameter represents marginal precision and
governs the total variability explained by the random component. The
mixing parameter distributes this variability between the spatially
structured and unstructured term. The given reparameterisation has a resemblance to
modelling the nugget effect in geostatistics, which has been
 explained as a sum of a true ``micro scale" nugget variance and
 variance due to measurement error \citep{cressie:93}.
By using the PC prior framework, we
are able to control the effect and informativeness of these random
fields using interpretable probability statements to tune the
hyperpriors in an intuitive way. Also, the scaling of the spatially
structured model ensures that given hyperprior choices can be
transferred for analyses using different grid resolutions.

The given model formulation represents a first step towards explicitly
communicating hyperprior choices in spatial point pattern analysis,
fitting a log-Gaussian Cox process to an observed spatial point
pattern. \.The framework of log-Gaussian Cox processes is a versatile
and flexible tool in spatial point pattern analysis, and care should
be taken to facilitate correct ecological interpretation of the
statistical results. The issues discussed here, 
are equally relevant in non-Bayesian settings where the choice of smoothing parameters impact the results in a similar way and an explicit discussion of these issues is equally ripe. An important future aim is to extend the given
ideas to more complex modelling of the log-intensity, for example
including non-linear effects of covariates, interaction between
species and also to joint models of point patterns with marks as in \cite{ho2008modelling} and \cite{illianal-a:12}. This
requires further work in terms of model formulation, in which the
total variance explained by random components is distributed between
several model parameters, aiming to make the effect and influence of all
hyperprior choices transparent.

\section*{Acknowledgements}
The authors wish to thank H\aa vard Rue for valuable discussions and for implementing 
the \texttt{rw2diid} model within the R-INLA framework. 

The BCI forest dynamics research project was founded by S.P. Hubbell
and R.B. Foster and is now managed by R. Condit, S. Lao, and R. Perez
under the Center for Tropical Forest Science and the Smithsonian
Tropical Research in Panama. Numerous organizations have provided
funding, principally the U.S. National Science Foundation, and
hundreds of field workers have contributed. Kriged estimates for
concentration of the soil nutrients were downloaded from 
http://ctfs.si.edu/webatlas/datasets/bci/soilmaps/BCIsoil.html. We
acknowledge the principal investigators that were responsible for
collecting and analysing the soil maps (Jim Dallin, Robert John, Kyle
Harms, Robert Stallard and Joe Yavitt), the funding sources (NSF
DEB021104,021115, 0212284,0212818 and OISE 0314581, STRI Soils
Initiative and CTFS) and field assistants (Paolo Segre and Juan Di
Trani).

\bibliographystyle{apa}
\bibliography{shs,local-bibfile}

\section{Appendix: A brief review on computation of penalised
    complexity priors} \label{appendix:pc}
    The
computation of PC priors for a given parameter $\xi$ is based on four
principles. For ease of readability, these are listed below. We refer
to \cite{simpson:17} for further details on computation of PC priors,
including a wide range of examples.
\begin{description}
\item[1. Occam' razor:] Assume that $f_1=\pi(x\mid \xi)$ represents a
    flexible version of a simpler base model
    $f_0=\pi(x\mid \xi=\xi_0)$. The prior for $\xi$ is computed to
    penalise deviation from the flexible model to the base model. This
    ensures that the constructed prior prefers the simpler model,
    until there is enough support for a more complex model.
\item[2. Measure of complexity:] The PC prior is assigned to a
    function of the Kullback-Leibler divergence (KLD)
    \citep{kullback:51}, which represents a measure of complexity
    between two densities. Explicitly,
    \begin{equation*}
        \mbox{KLD}(f_1 \parallel f_0)=\int f_1(x)\log\left(\frac{f_1(x)}{f_0(x)}\right)dx,\label{eq:kl}
    \end{equation*} 
    To facilitate interpretation, this deviation is transformed to a
    unidirectional measure of distance between the two densities,
    defined by
    \begin{equation*}
        d(f_1\parallel f_0)= \sqrt{2\mbox{KLD}(f_1 \parallel f_0)}. \label{eq:dist}
    \end{equation*}
\item[3. Constant-rate penalisation:] The prior for $d$ is chosen
    according to a principle of constant rate penalisation
    \begin{equation*}
        \frac{\pi(d+\delta)}{\pi(d)}=r^{\delta},\quad d,\delta\geq 0,\label{eq:const-rate}
    \end{equation*}
    where $r\in (0,1)$. Consequently, the relative change in the prior
    for $d$ is independent of the actual distance and $d$ is
    exponentially distributed,
$$\pi(d)=\lambda\exp(-\lambda d),$$ where $r=\exp(\lambda)$.  The prior for $\xi$  follows by a standard change of variable transformation and will always have its mode at the base model. 
\item[4. User-defined scaling:] The effect (or size of deviation from
    the base model) of a random component is adjusted by an intuitive
    scaling parameter, incorporating a user-defined probability
    statement for the parameter of interest. For example, this
    statement can be defined in terms of tail events, for
    example $$P(Q(\xi)>U)=\alpha,$$ where $U$ represents an assumed
    upper limit for an interpretable transformation $Q(\xi)$, while
    $\alpha$ is a small probability, see on PC priors. This implies
    that the user should have an idea of a sensible size of the
    parameter (or any transformed variation) of interest.
\end{description}

\end{document}